\documentclass[a4paper,twocolumn,10pt]{quantumarticle}
\pdfoutput=1

\usepackage[utf8]{inputenc}
\usepackage[english]{babel}
\usepackage[T1]{fontenc}
\usepackage{graphicx}
\usepackage{dcolumn}
\usepackage{titletoc}
\usepackage{bm}
\usepackage{xcolor}
\usepackage{physics}
\usepackage{quantikz} 
\usepackage{soul}
\usepackage{enumitem}
\usepackage{comment}
\usepackage[normalem]{ulem}
\usepackage{tabularx}
\usepackage{booktabs}

\usepackage{amsmath, amssymb, amsthm}
\theoremstyle{plain}
\newtheorem{theorem}{Theorem}[section]     
\newtheorem{proposition}[theorem]{Proposition} 
\newtheorem{lemma}[theorem]{Lemma}             
\newtheorem{conjecture}[theorem]{Conjecture}   
\theoremstyle{definition}
\newtheorem{definition}[theorem]{Definition}   

\theoremstyle{remark}
\newtheorem*{evidence}{Evidence} 
\newtheorem*{heuristic}{Heuristic}

\usepackage[numbers,sort&compress]{natbib}

\definecolor{dark-red}{rgb}{0.84,0.15,0.15}
\definecolor{dark-blue}{rgb}{0.15,0.15,0.8}
\definecolor{medium-blue}{rgb}{0,0,0.5}
\definecolor{copper}{rgb}{0.72, 0.45, 0.2}

\usepackage{hyperref}
\hypersetup{
    colorlinks, linkcolor={dark-red},
    citecolor={dark-blue}, urlcolor={medium-blue}
}

\usepackage[capitalize,nameinlink]{cleveref}
\definecolor{copper}{rgb}{0.72, 0.45, 0.2}
\definecolor{dark-green}{rgb}{0.13, 0.55, 0.13}

\newcommand{\superket}[1]{{\left|#1\rangle \rangle \right.}}
\newcommand{\pcmiltrange}{p_c^{\text{MILT}} \approx 0.10\text{--}0.15}

\usepackage{silence}
\makeatletter
\providecommand{\@afterenddocumenthook}{}
\makeatother

\begin{document}

\title{Measurement-Induced Landscape Transitions and Coding Barren Plateaus in Variational Quantum Circuits}

\author{Gaurav Gyawali$^{\dagger,}$}
\email{gaurav.gyawali@hpe.com}
\thanks{Current address: HPE Quantum, Hewlett Packard Enterprise Labs, California, USA}
\affiliation{Laboratory of Atomic and Solid State Physics, Cornell University, Ithaca, NY 14853, USA}
\affiliation{Department of Physics, Harvard University, Cambridge, MA 02138, USA}

\author{Sonny Rappaport$^{\dagger,}$}
\affiliation{Laboratory of Atomic and Solid State Physics, Cornell University, Ithaca, NY 14853, USA}

\author{Tiago Sereno}
\affiliation{Department of Physics, Applied Physics, and Astronomy, Binghamton University, Binghamton, NY 13902, USA}
\affiliation{School of Physics and Astronomy, University of Minnesota, Minneapolis, MN 55455, USA}

\author{Michael J. Lawler}
\email{mlawler@binghamton.edu}
\affiliation{Department of Physics, Applied Physics, and Astronomy, Binghamton University, Binghamton, NY 13902, USA}
\affiliation{Laboratory of Atomic and Solid State Physics, Cornell University, Ithaca, NY 14853, USA}
\affiliation{Department of Physics, Harvard University, Cambridge, MA 02138, USA}

\begin{abstract}
Barren Plateaus---the exponential vanishing of parameter gradients with system size---limits the practical application of variational quantum algorithms (VQA). Recently, a ``landscape transition'' from a barren to a navigable plateau was observed in monitored quantum circuits, hypothesized to coincide with the measurement-induced phase transition (MIPT) separating area-law from the volume-law states. 
However, from an information-theoretic standpoint, we argue these transitions are fundamentally distinct. We support this hypothesis with numerical and analytical evidence---including quantum-classical channel mutual information measures, visualizations of the optimization trajectories, and an analytical statistical mechanics model of cost-gradient variances.
Our results provide evidence for a measurement-induced landscape transition (MILT) at $\pcmiltrange < p_c^{\text{MIPT}}$ out of a barren plateau phase with rigorous analytical support if the measurements take place far from the varied parameter associated with the gradient. Furthermore,  throughout $0<p<p_c^{\text{MILT}}$, we find a finite mutual information between Alice's information encoded in the parameters and Bob's information extracted from final time measurements. These results point toward barren plateau mitigation via optimization algorithm design guided by the mutual information and through the strategic incorporation of mid-circuit measurements.
\end{abstract}

\maketitle
\def\thefootnote{$\dagger$}\footnotetext{These authors contributed equally to this work.}\def\thefootnote{\arabic{footnote}}

\section{Introduction}

Variational quantum algorithms (VQAs) are considered a promising approach to solving a broad range of problems using quantum computers \cite{Peruzzo_VQA_2014,McClean_VQA_2016, Cerezo_VQA_2021, Bharti_2022_NISQ}. At their core, VQAs map computational problems to optimization tasks, employing a hybrid classical-quantum strategy to iteratively update parameterized quantum circuits and minimize a defined cost function. In principle, these algorithms can prepare quantum states that lie beyond the reach of classical simulation methods, including state-of-the-art tensor network techniques \cite{haghshenas2022variational}. Yet, this theoretical promise has thus far gone largely unrealized. The central obstacle is the barren plateau (BP) phenomenon—the exponential vanishing of cost function gradients with system size—which renders gradient-based optimization infeasible before the classically hard regime is reached \cite{McClean_barren_plateaus_2018,Larocca_2025_barren_plateaus_in_variational_quantum_computing,Wang_noise_induced_bp, Marrero_entanglement_bp_2021}. Solving the BP problem is therefore a key to unlocking the practical quantum advantage that VQAs were designed to deliver.

Numerous strategies have been proposed to mitigate BPs. These include problem-inspired \cite{Mele_hamiltonian_variational_ansatz_2022} and adaptive ansätze \cite{Grimsley_adaptive_bp_2023}, local cost functions \cite{Cerezo_cost_function_bp_2021}, pre-training techniques \cite{Friedrich_pre_training_2022}, layerwise learning \cite{Skolik_2021_layerwise_learning}, and identity-block initialization \cite{Grant_initialization_2019}. Additionally, recent studies have explored the critical relationship between entanglement growth in random circuits and the onset of BPs \cite{Marrero_entanglement_bp_2021, Patti_entanglement_bp_2021, Sack_2022}. For example, computing many-body entanglement during the training process can help avoid BPs by dynamically controlling the learning rate \cite{Sack_2022}. Nevertheless, existing mitigation strategies typically exchange BPs for other significant bottlenecks such as increased circuit depths, substantial computational overhead, or restricted ansatz expressibility. Consequently, a scalable solution remains elusive.

Since entanglement is intimately connected to the onset of BPs \cite{Marrero_entanglement_bp_2021, Patti_entanglement_bp_2021}, a natural question is whether controlling entanglement through mid-circuit measurements can offer a route out of the barren plateau. Without measurements, the entanglement entropy of a quantum chain undergoing local unitary evolution grows linearly in time until it saturates to a volume-law entangled state \cite{Nahum_entanglement_growth_2017}. Local measurements counteract this growth via wavefunction collapse, and a dynamical competition between the two drives a phase transition between phases with different system-size scaling of entanglement — the measurement-induced phase transition (MIPT) \cite{Li_2018_quantum_Zeno_MIPT, Li_MIPT_2019, Skinner_MIPT_2019, Jian_measurement_induced_criticality_2020, Lavasani_monitored_2022, Lavasani_symmetric_2022, Choi_QEC, Gullans_dynamical_purification_2020}. Indeed, it was recently observed that mid-circuit measurements can drive a "landscape transition" in variational circuits from BP to mild or no BP \cite{wiersema2023measurement}, with numerical results suggesting this transition coincides with the MIPT and exhibits similar critical exponents. Owing to the close relationship between entanglement and BPs, it is tempting to conclude that the landscape transition and the entanglement transition belong to the same universality class — but, as we argue below, this picture is incomplete.

In this work, we study VQAs in the presence of projective measurements interspersed with unitary layers, considering two cases: the \textit{mixed} case, in which measurement outcomes are discarded, and the \textit{post-selection} case, in which they are retained. Our results---supported by landscape visualizations, information-theoretic measures, and a rigorous statistical mechanics model---demonstrate that both cases undergo a measurement-induced landscape transition (MILT) at $p_c^{\text{MILT}} \approx 0.1$--$0.15$, well below the entanglement transition $p_c^{\text{MIPT}}$. We argue from an information-theoretic perspective that the MILT and MIPT are transitions in fundamentally different quantities: the mutual information governing classical channel capacity and the coherent information governing quantum channel capacity, respectively. From statistical mechanics perspective, they are described by two different partition functions with distinct boundary conditions for the domain-wall dynamics leading to configurations with different topology. 

Furthermore, as we increase the probability of measurement from 0 to $p_c^{\textrm{MILT}}$, the barren plateau gets less severe, and this finding is supported by the landscape visualization which shows broadening of of the valley around the minimum. The mutual information reveals that the phase $0 < p < p_c^{\text{MILT}}$ constitutes a \textit{coding barren plateau}: a regime in which information about the circuit parameters reaches the local cost function through the optimization channel even though the gradient vanishes. At low measurement rate, post-selection may not a bottleneck in this phase, suggesting a path toward scalable measurement-assisted optimization.

\subsection{Information theory of optimization}
\label{sec:infotheory}

\begin{figure}[t]
    \centering
    \includegraphics[width=0.95\columnwidth]{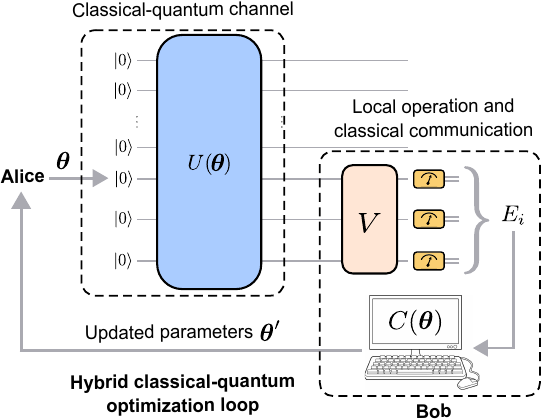}
    \caption{
    \textbf{An information-theoretic perspective on hybrid classical-quantum optimization.} The optimization loop is cast as a two-step communication protocol between two parties. (\textit{Step 1}) Alice encodes parameters $\boldsymbol \theta$ into a quantum state via a classical-quantum channel $\boldsymbol{\theta} \mapsto \rho_{\boldsymbol{\theta}}$. (\textit{Step 2}) Bob performs a POVM measurement $\{E_i\}$ on $\rho_{\boldsymbol{\theta}}$ to computes the cost function $C(\boldsymbol{\theta})$ and the updated parameters $\boldsymbol \theta'$ that he sends to Alice via a classical channel. Bob's action is restricted to local operations and classical communication (LOCC). The capacity of this end-to-end channel to transmit information about $\boldsymbol{\theta}$ determines the success of optimization and forms the basis for our information-theoretic analysis of the MILT throughout this work.
    }
    \label{fig:fig1}
\end{figure}

We can view VQAs as an information-processing task in which classical parameters are encoded into quantum states. Consider a parameterized ansatz $|\psi(\boldsymbol{\theta})\rangle = U(\boldsymbol{\theta})|0\rangle^{\otimes N}$, where the circuit $U(\boldsymbol{\theta}) = U_L(\theta_L) \cdots U_1(\theta_1)$ is composed of $L$ layers with $U_k = e^{i\theta_k G_k}$ generated by Hermitian operators $\{G_k\}$. The cost function is evaluated by measuring an observable $O$,
\begin{equation}
C(\boldsymbol{\theta}) = \langle\psi(\boldsymbol{\theta})|O|\psi(\boldsymbol{\theta})\rangle = \langle 0|U^\dagger(\boldsymbol{\theta})OU(\boldsymbol{\theta})|0\rangle,
\end{equation}
and minimized over $\boldsymbol{\theta}$ by a classical optimizer.

This standard formulation can be recast as a communication problem, as illustrated in \cref{fig:fig1}. Alice sends classical information to Bob by encoding parameters $\boldsymbol\theta$ into a quantum state via a classical-quantum channel  $\mathcal{N}: \boldsymbol{\theta} \mapsto \rho_{\boldsymbol{\theta}}$ (see Chapter 5 of \citeauthor{holevo2019quantum}~\cite{holevo2019quantum}). Bob has local access to the quantum states via his register, and he attempts to decode Alice's parameters using POVMs $\{E_i\}$ associated with the eigenbasis of $O$. Restricting Bob to a single projective basis reduces the channel to the classical conditional distribution 
\begin{equation}
p(i|\boldsymbol{\theta}) = \mathrm{Tr}[\rho_{\boldsymbol{\theta}}E_i].
\end{equation}
Bob uses the measurement outcomes to evaluate the cost function $C(\boldsymbol \theta)$, which is returned to Alice via a noiseless classical channel---a sequence of steps constituting local operations and classical communication (LOCC).

Before proceeding, we note an important distinction between the communication analogy and the actual optimization task. In the standard communication setting, Alice draws parameters from a fixed prior $p(\boldsymbol{\theta})$ with the goal of transmitting them faithfully to Bob. In VQA optimization, Alice does not draw from a fixed prior except at initialization; she iteratively updates $\boldsymbol{\theta}$ based on the cost function values returned by Bob, following a gradient descent or similar rule. The optimization loop therefore does not directly measure the channel capacity. Nevertheless, the capacity of the quantum channel to transmit classical information about $\boldsymbol{\theta}$ remains the relevant figure of merit: if Bob cannot distinguish outputs corresponding to different parameters, no optimizer---gradient-based or otherwise---can extract useful information from the cost landscape. Quantifying this capacity therefore provides a rigorous upper bound on trainability, a connection we now make precise.

Communicating classical parameters through a quantum channel is a foundational problem in quantum Shannon theory~\cite{wilde_2013_qit}, rigorously framed by the Holevo-Schumacher-Westmoreland (HSW) theorem~\cite{schumacher_1997_sending_classical_information}. While the theorem formally defines capacity via a regularization over infinite channel uses, the trainability of a single variational circuit is governed by the single-use Holevo information of the classical-quantum channel i.e., $\chi$. This quantity corresponds to the maximum mutual information between Alice's classical information and Bob's quantum register. Given Alice's distribution $p(\boldsymbol{\theta})$, it is defined as

\begin{equation}
    \chi(p(\boldsymbol{\theta})) \equiv  S \left(\sum_{\boldsymbol{\theta}} p(\boldsymbol{\theta}) \rho_{\boldsymbol{\theta}}\right) - \sum_{\boldsymbol{\theta}} p(\boldsymbol{\theta}) S\left( \rho_{\boldsymbol{\theta}}\right),
\end{equation}
where, $S(\rho)$ is the von Neumann entropy $S( \rho) = -\Tr[\rho \log \rho]$.

While the HSW theorem assumes the ability to perform collective measurements over many copies, here, Bob is restricted to a fixed, local measurement basis (determined by the cost observable $O$) on single copies, though one could search through different observables to identify optimizable problems. Thus, the mutual information should be computed between Alice's classical probability distribution and Bob's measurement outcomes. The calculation details for this classical mutual information are given in  \cref{sec:IAB_methods}. We choose a uniform probability distribution for Alice for comparison against the gradient variance results which are usually computed for uniformly drawn parameters. The maximum amount of information that can be extracted by Bob is quantified by the accessible mutual information, 
\begin{equation}
    \label{eqn:I_acc}
    I_{\mathrm{acc}}  \equiv \max_{p(\boldsymbol \theta), \: \{E_i\}} I(A:B),
\end{equation}
where the maximization is performed over both Alice's distribution as well as Bob's POVMs $\{E_i\}$. By definition, the Holevo quantity upper bounds $I_{\mathrm{acc}}(\mathcal{N})$, i.e., $I_{\text{acc}}\le \chi$ \cite{wilde_2013_qit}. Thus, $\chi(\mathcal{N})$ serves as a fundamental upper bound on trainability; if $\chi(\mathcal{N})$ vanishes, no local measurement strategy can recover the parameters.

If any intermediate measurements occur in the circuit, we consider two scenarios depending on Bob's access to the data. If the measurement outcomes $\mathbf{M}$ are available to Bob (the post-selected setting), the classical-quantum channel becomes ${\boldsymbol\theta}\mapsto\ \rho_{{\boldsymbol\theta},\mathbf{M}}$, where $\rho_{{\boldsymbol\theta},\mathbf{M}}$ is the pure state resulting from the collapse of the wavefunction. Conversely, if the measurement outcomes are discarded or unknown to Bob (the mixed setting), the effective channel averages over all possible measurement trajectories. This averaging acts as a strong decoherence, driving the effective state $\rho_{\boldsymbol\theta}$ toward the maximally mixed state $\mathbb{I}/d$. In this work, we refer to this noise-induced approach to the maximally mixed state as ``thermalization.''

We can now understand the landscape transition by viewing optimization as a communication problem with two distinct failure modes. The first mode occurs when there are no measurement gates ($p=0$). Here, the deep circuit scrambles the information so thoroughly that the dependence of the state on the parameters $\boldsymbol{\theta}$ is hidden from local observables into the exponentially large Hilbert space (the BP). The second mode occurs when measurements are frequent ($p \to 1$). In this ``Quantum Zeno'' regime, the state is projected so frequently that the dynamics freezes, effectively decoupling the state from the unitary parameters $\boldsymbol{\theta}$. If Bob cannot access the outcomes, this frequent projection drives the rapid thermalization described above, destroying the signal. This destruction, though, is system size independent and distinct from the first failure mode. We find that a dynamic competition between either of these two modes drives a landscape transition observed in the cost-gradient variance. Crucially, the success of optimization is governed by the capacity of the end-to-end channel shown in \cref{fig:fig1} to transmit classical information.

\subsection{Variational ansatzes}
\label{sec:ansatzes}

\begin{figure*}[th!]
    \includegraphics[width=0.9\textwidth]{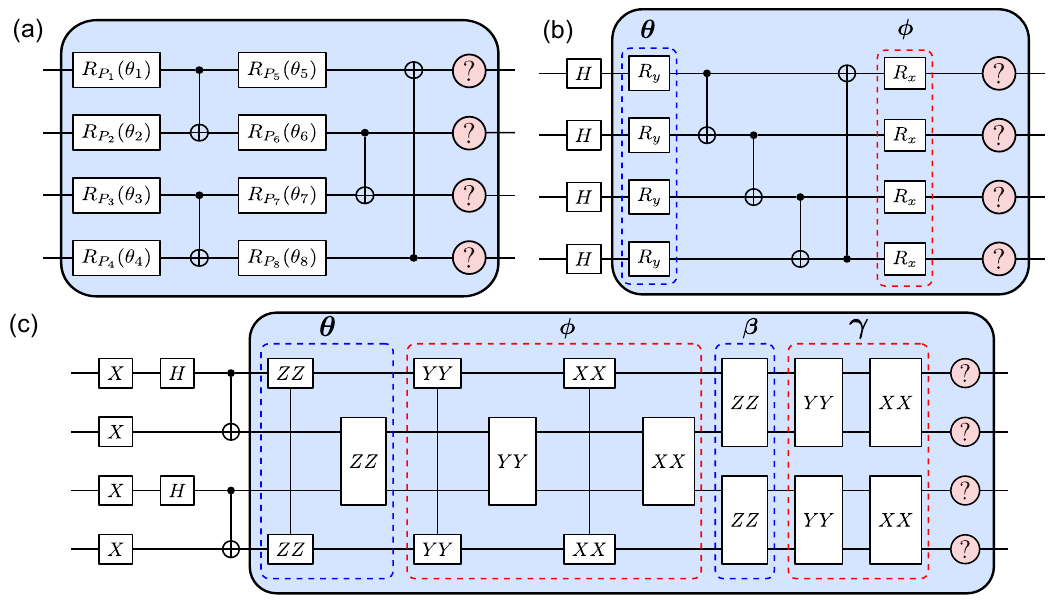}
    \caption{\textbf{A single layer of the ansatzes examined in this paper.} The light-orange shaded regions in each diagram are those repeated in every layer, whereas those outside are used only for the initial layer. Measurement gates, represented by the light-red circles, are randomly placed with probability $p$ after each layer except the last. \textbf{a} HEA1: Apply a $R_{P_i}$ gate to each qubit, followed by CNOT pairs on even paired qubits, followed by applying more $R_{P_i}$ gates to each qubit, followed by applying CNOT pairs to odd paired qubits. Each $R_{P_i}$ gate represents a random rotation gate generated by a Pauli operator $P_i$, parameterized by an angle ${\theta}_i$. \textbf{b} HEA2 and \textbf{c} XXZ-HVA are heavily inspired by the ansatzes used in  \citeauthor{wiersema2023measurement}  \cite{wiersema2023measurement}. Note that the parameters are applied uniformly across all the qubits in \textbf{b} and \textbf{c}. Although shown for only four qubits, each ansatz can be generalized to any even number of qubits.}
    \label{fig:ansatzes}
\end{figure*}

Finding a desired ground state within an exponentially large Hilbert space with VQA is challenging due to the competition between \emph{expressibility} and \emph{trainability}. An ideal variational ansatz must be highly expressible \cite{Sim_expressibility_2019,Haug_2021_expressibility_quantum_geometry, Holmes_expressibility_2022}, i.e., it should be capable of creating a complex network of entanglement between qubits. Yet, it 
should remain trainable---restricting the search to a polynomial-sized subspace relevant to the solution that a classical optimizer can navigate \cite{ragone_2024_lie_algebraic_theory}. However, these two demands are in direct tension: high expressibility generates highly entangled states and a landscape plagued by BPs with narrow gorges \cite{Cerezo_cost_function_bp_2021}, rendering gradient-based optimization challenging.

The first step in addressing this challenge is to carefully design the variational ansatz. We choose several ansatzes to capture the typical behavior of landscape transitions and their information-theoretic properties. 
The fundamental building blocks of a variational ansatz include arbitrary single-qubit rotations and entangling operations, such as CNOT and CZ gates. A hardware-efficient ansatz (HEA) is constructed by selecting two-qubit gates and a connectivity topology that correspond to the underlying device architecture \cite{Kandala_HEA_2017}. Although flexible and highly expressible, optimizing them can be challenging. Thus, it is often useful to turn to problem-inspired ansatzes, such as the Quantum Approximate Optimization Algorithm (QAOA)~\cite{Farhi_2014_QAOA} for combinatorial problems and the Hamiltonian Variational Ansatz (HVA)~\cite{Wecker_progress_towards_2015,Katemolle_2022_kagome_vqe} for finding ground states. For specific problems, such as learning phase diagrams, renormalization group-like flows can be implemented using Quantum Convolutional Neural Networks (QCNNs)~\cite{Cong_QCNN_2019}, which have been found to be free from BPs~\cite{Pesah_2021_absence_of_bp}. Alternatively, one may optimize how circuit blocks are constructed dynamically. An example of this paradigm is the ADAPT-VQE algorithm~\cite{Grimsley_2019_ADAPT,GG_2022_ADAPT}, which is also notably BP-free~\cite{Grimsley_adaptive_bp_2023}. Motivated by these advances, our work seeks to enhance standard ansatzes by introducing measurement gates to explicitly control entanglement growth.

We focus specifically on the following four ansatzes:
\begin{enumerate}
    \item[(a)] Hardware efficient ansatz 1~(HEA1)
    \item[(b)] Hardware efficient ansatz 2~(HEA2)
    \item[(c)] XXZ Hamiltonian variational ansatz~(XXZ-HVA)
    \item[(d)] Haar unitary ansatz (HUA)
\end{enumerate}
The first three ansatzes chosen are similar to those used in the paper by  \citeauthor{wiersema2023measurement}   \cite{wiersema2023measurement}, in part to provide a direct comparison of our results. Specifically, our HEA2 and XXZ-HVA ansatz are nearly identical to their HEA and XXZ-HVA. The fourth ansatz is the brickwork Haar unitary ansatz consisting of arbitrary 2-qubit unitary gates. All ansatzes consist of several layers, in which unitary gates are applied. In this work, we focus on a hybrid setting where the unitary layers are followed by projective measurements with probability $p$ on each qubit, a setup commonly used in the MIPT literature. Ansatzes (a), (b), and (c) are shown in \cref{fig:ansatzes}.

\section{Results}
We first turn to our numerical results on HEA1, HEA2, and XXZ-HVA and then follow up with analytical insight from HUA. In what follows, we provide evidence the MILT and MIPT are different phase transitions, and there exists a coding BP phase in the range $0<p<p^{\text{MILT}}_c$. We will do so by observing the variances of cost function gradients, visualizations of the optimization process, and information-theoretic measures of the optimization process.

\subsection{Landscape transitions observed in mixed and projective gradients}
\label{sec:variance_study}
\begin{figure*}[ht]
\begin{center}
\includegraphics[width=0.9\textwidth]{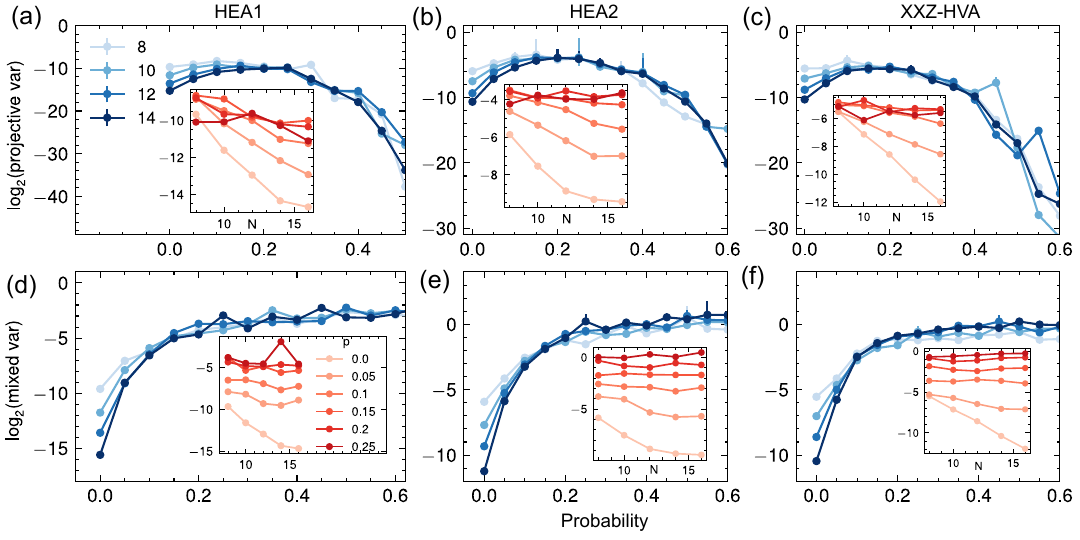}
\caption{\textbf{Log of the variance for projective and mixed gradients.} Top panels ((a), (b), (c)) show the projective gradient's variance vs. measurement probability for HEA1, HEA2, and XXZ-HVA ansatzes respectively for system sizes $N=8-14$ at constant depth $16$. Inset shows the projective variance as a function of the number of qubits ($N$), showing that the BP becomes less severe with higher measurement probability for $p \approx 0.1-0.15$. Nevertheless, the projective gradient decays exponentially with probability after a certain value. Similarly, the bottom panels ((d), (e), (f)) show the mixed gradient's variance for  HEA1, HEA2, and XXZ-HVA ansatzes, respectively. Insets show the variance as a function of the number of qubits, showing the BP becomes less severe with more measurement gates and completely disappears around $\pcmiltrange$ depending on ansatz.}
\label{fig:projective_gradients}
\end{center}
\end{figure*}

\begin{figure*}[th]
    \includegraphics[width=0.9\textwidth]{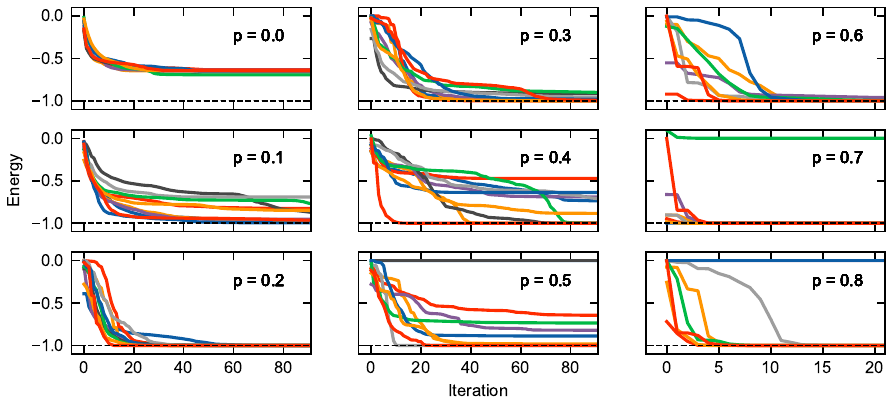}
    \caption{\textbf{Optimization with measurement and post-selection.}  Optimization traces shown for a 8-qubit system with cost function $Z_0Z_1$. Each sub-figure corresponding to a measurement probability between 0 and 0.8 contains traces from 10 individual optimization runs carried out for the HEA2 ansatz with depth 20. The dashed line shows the exact ground state energy.}
    \label{fig:z0z1_optimization}
\end{figure*}

To characterize the trainability of the ansatzes depicted in \cref{fig:ansatzes}, we computed the variance of the cost function gradients by randomly sampling parameters from a uniform distribution, the standard diagnostic for BPs~\cite{McClean_barren_plateaus_2018}, across system sizes up to $N=18$ and circuit depths up to $L=100$. Before discussing these results, we carefully distinguish between the two types of cost functions that arise from monitored dynamics.

The intermediate measurements in hybrid variational quantum circuits yield different outcomes for each run. We can compute the expectation value of observables either by (i) post-selecting for a specific outcome set $\mathbf{M}$ or (ii) averaging over all the measurement outcomes. Post-selection requires exponentially many samples in the number of measurement gates. However, if the expectation value is averaged over all the measurement outcomes, it corresponds to computing the expectation value with respect to the average density matrix. Thus, we define two types of cost functions, \emph{projective} cost functions computed with post-selected measurement outcomes, and \emph{mixed} cost functions computed ignoring measurement outcomes. 

When mid-circuit measurements are introduced, the circuit's evolution becomes stochastic. For a given sequence of measurement outcomes $\mathbf{M}$, the (unnormalized) final state is $|\tilde{\psi}_\mathbf{M}\rangle = \left( \prod_{k=L}^1 P_{\mathbf{M}_k} U_k(\theta_k) \right) |0\rangle$, where $P_{\mathbf{M}_k}$ is the projector for outcome at layer $k$. The probability of this outcome sequence is $p_\mathbf{M}(\boldsymbol{\theta}) = \langle \tilde{\psi}_\mathbf{M} | \tilde{\psi}_\mathbf{M} \rangle$. Such final states enable us to define two distinct cost functions.

\emph{The projective cost function} refers to the expectation value of an observable $O$ calculated for a specific post-selected wave function $|\psi_{\mathbf M}\rangle = |\tilde\psi_{\mathbf M}\rangle/\sqrt{p_{\mathbf M}}$. We define it as 
\begin{equation}
    C_{\mathbf M}(\boldsymbol{\theta}) = \expval{O}{\psi_\mathbf{M}({\boldsymbol\theta})}.
\end{equation}
Likewise, \emph{the mixed cost function} is obtained by averaging the expectation value of an observable $O$ over all possible measurement outcomes i.e.,
\begin{align}
    C({\boldsymbol\theta}) = \mathbb{E}[ C_\mathbf{M}({\boldsymbol\theta})] &= \sum_{\mathbf{M}} p_{\mathbf{M}}({\boldsymbol\theta}) \expval{O}{\psi_\mathbf{M}({\boldsymbol\theta})} \nonumber\\
    &= \Tr[O \rho({\boldsymbol\theta})]. 
\end{align}
In other words, the mixed cost function is the expectation value with respect to the measurement-averaged density matrix $\rho = \sum_\mathbf{M} p_\mathbf{M} \rho_\mathbf{M}$. For conciseness, we call the gradients of these two cost functions \emph{projective} and \emph{mixed gradient} respectively. A more detailed explanation of how these cost functions and their gradients are computed is presented in \cref{app:computing_gradients}.

In \cref{fig:projective_gradients}, we show the variance of the projective cost function (top panels) for three ansatzes---HEA1, HEA2 and XXZ-HVA---as a function of measurement probability $p$. For all system sizes $N$, we report the variance at a constant depth $L=16$. At $p=0$, we find that $L=16$ is already sufficient for all reported system sizes to exhibit a barren plateau, as evidenced by the exponential decay with $N$ shown in the inset (lightest red curve). As $p$ increases, the barren plateau becomes less severe, with the variance curves for all system sizes eventually crossing at an ansatz-dependent critical value $p_c^{\text{MILT}}$, in qualitative agreement with Ref.~\cite{wiersema2023measurement}. After this threshold, the variance stops decaying with the system size but instead decays with probability independent of the system size until it hits the machine precision. The location of this transition in $p$ is examined in more detail in \cref{app:critical_point_analysis}.

It is important to note that projective gradients in \cref{fig:projective_gradients} are averaged over various measurement outcomes, so the error bars are higher for larger $p$ as the number of possible outcomes scales like $2^M$, where $M$ is the number of measurement gates. We refer to this region of exponentially small gradients as the ``quantum Zeno" phase, a term coined by  \citeauthor{Li_2018_quantum_Zeno_MIPT}  \cite{Li_2018_quantum_Zeno_MIPT} for quantum states that are frequently measured, and hence stalled close to an eigenstate of the measurement operator. Moreover, we are post-selecting for outcomes $\mathbf{M}$; consider $\mathbf{M} = \{0,0\cdots, 0 \}$ and $p=1.0$, then the expectation value of an operator $O$ is always going to be $\expval{O}{00\cdots00}$ with no dependence on the variational parameters. The measurements essentially cut the communication between the parameters at the beginning of the circuit and the end, hence the diminishing gradients. 

The value of probability at which the BP disappears depends on two factors: (i) the depth of the circuit and (ii) ansatz. The effect of depth on the projective gradients is to shift the turning point towards $p=0$, eventually stopping at $\pcmiltrange$. The dependence on the ansatz is much weaker than the dependence on the circuit depth. As a result, we do not find the landscape transition coinciding with the MIPT in our projective gradients as reported by  \citeauthor{wiersema2023measurement}  \cite{wiersema2023measurement} (see \cref{subsec:mutual_info_results} for details).

The mixed gradients always seem to have the same basic structure i.e., the gradient grows with the probability of measurement and plateaus to a constant value after around $\pcmiltrange$. Exponential vanishing of gradients i.e, a BP, in this case is compensated for by summing over exponentially many measurements. Here, we still destroy the BP at the same critical value as in the case of projective gradients but this does not  seem useful for optimization because it introduces thermalization.

\subsection{Landscape visualization of the optimization process}

The cost gradient variance averaged over uniformly random parameters analyzed in the previous section is a standard diagnostic for barren plateaus, but it can faithfully characterize the landscape only at initialization.  As optimization progresses, parameters concentrate near regions of low cost, and the relevant landscape geometry may differ substantially from its random-parameter average. Moreover, gradient variance is a second-moment quantity: it cannot resolve directional structure or curvature in the landscape that adaptive optimizers could in principle exploit \cite{Stokes_2020_quantum_natural_gradient, Haug_2021_expressibility_quantum_geometry}. It is also uninformative for gradient-free strategies, such as Bayesian optimization \cite{Shahriari_2016_bayesian_optimization} and sequential minimal optimization \cite{Nakanishi_2020_sequential_minimal}, which navigate the landscape by modeling its global geometry rather than following local gradient information. To address these limitations, we complement the gradient analysis with optimization traces that track the algorithm's path through the cost landscape, and with a direct visualization of the landscape geometry near a local minimum using the technique of \citeauthor{li2018visualizing} \cite{li2018visualizing}

In short, we find that the optimization traces benefit from the improved variance of hybrid variational ansatzes, thus mitigating the BP problem. The optimization runs were carried out for the projective cost function corresponding to the HEA2 ansatz for $Z_0 Z_1$ and $XXZ$ Hamiltonians. For the $Z_0 Z_1$ Hamiltonian, the cost function depends only on the first two qubits, and the initial computational basis state satisfies the minimum value. Nevertheless, optimizing a circuit with a depth of 20 to get this simple product state is challenging because of the BP, as shown in \cref{fig:z0z1_optimization}. Inserting measurement gates results in the optimization traces exploring various energies, with some reaching the minima. For higher probabilities, the optimization traces again get stuck at some parameter values. However, most happen to satisfy the minimum value because of the simple nature of the cost function with enormously degenerate global minima.  

In \cref{app:xxz_traces}, we present optimization traces on the $XXZ$ Hamiltonian ground state problem, which show similar but more complex behavior. In general, we find measurements seem to help for some ansatzes more than others. For the HEA2 ansatz, they seem to work well, but for HEA2 with flexible parameters (not shown), and the XXZ-HVA the results are more mixed.

Although the variance of the mixed gradient increases with an increase in the number of measurement gates, they optimize poorly. The optimization trace (not shown here) gets noisier with the probability of placing a measurement. This suggests a noisy cost landscape for the mixed cost function.

\begin{figure}[ht]
    \centering
    \includegraphics[width=0.95\columnwidth]{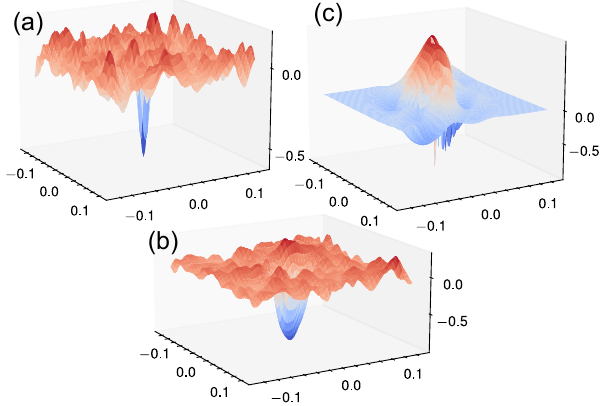}
    \caption{\textbf{Cost landscape visualizations of the monitored HEA2 hybrid variational circuits with $O = Z_0Z_1$.} Visualizations are produced by optimizing the circuit and then plotting $C({\boldsymbol\theta})$ along two random directions in ${\boldsymbol\theta}$-space, where the origin is the minima achieved. (a) no measurements showing a large BP with a narrow path to the optimal state, (b) measurement probability $p=0.1$ showing a small BP with a broad path to the optimal state, and  (c) $p=0.8$ showing the quantum Zeno region, where the landscape is smooth and flat with broad suboptimal minima and a large volcano-like entrance to the optimal state.}
    \label{fig::loss_landscape}
\end{figure}

Since the optimization traces are complex, we turned to visualizing the cost landscape near a local minimum. Following Ref.  \cite{li2018visualizing}, we can visualize the cost landscape by randomly choosing two directions in Alice's parameter space and plotting the cost function on the plane spanned by these essentially orthogonal directions. The results for the HEA2 ansatz with the $Z_0 Z_1$ cost function are presented in  \cref{fig::loss_landscape}. In the absence of measurements, they show a rough BP with a very sharp drop to the local minima. This sharp drop broadens in the presence of measurements while the BP nevertheless persists. Then at a large measurement rate, in the quantum Zeno phase, the BP is absent, the landscape is smooth, and a volcano-like entrance guards the path to the minimum, explaining the difficulty of finding this minimum. 

Together, the gradient variance, optimization traces, and landscape visualizations establish the phenomenology of the MILT, but they do not by themselves reveal whether this transition is fundamentally distinct from the MIPT. We address this question next through an information-theoretic analysis.

\subsection{Differing information-theoretic measures of MIPT and MILT}
\label{subsec:mutual_info_results}

We now contrast the information-theoretic signatures of the MILT and MIPT to establish that they are fundamentally different transitions. The MIPT is a phase transition in the entanglement structure of monitored random quantum circuits, driven by the competition between scrambling unitary dynamics and localizing projective measurements. When measurements are sparse ($p < p_c^{\text{MIPT}}$), the late-time entanglement entropy of a subsystem $A$ exhibits volume-law scaling, $S(A) \propto |A|$. When measurements are frequent ($p > p_c^{\text{MIPT}}$), the state is repeatedly projected toward a product state and the entanglement follows an area law, $S(A) \propto |\partial A|$. At the critical point $p = p_c^{\text{MIPT}}$, the entanglement exhibits a logarithmic violation of the area law, $S(A) \propto \ln|A|$, characteristic of a conformal field theory \cite{Li_2018_quantum_Zeno_MIPT, Skinner_MIPT_2019, Choi_QEC}. Crucially, this transition occurs in the entanglement entropy, an observable non-linear in the reduced density matrix, and cannot be detected by any linear observable or its classical average.

In contrast, the partial derivative with respect to a parameter $\theta$ is a linear observable, defined below for a circuit $U =U_1(\theta)U_2$:
\begin{align}
\pdv{\expval{O}}{\theta} = \pdv{}{\theta} \expval{U_1^\dagger(\theta) U_2^\dagger O U_2 U_1(\theta)}\nonumber\\
= i\langle 0|U_1^\dagger(\theta) [G,U^\dagger_2 O U_2] U_1(\theta)|0\rangle,
\end{align}
where we have taken $U_1 = e^{-i\theta G}$, and $U_2$ is an arbitrary unitary  that follows $U_1$.  The variance of the linear observable is a classical average of this linear observable squared. MIPT is an entanglement transition i.e., a transition in an observable non-linear in the density matrix,  that can neither be observed in linear observables nor their classical averages. Therefore, it follows that $p^{\text{MILT}}_c \neq p^{\text{MIPT}}_c$.

From the perspective of quantum communication, the MILT and MIPT are transitions in two distinct information-theoretic quantities. In the MIPT, the competition between unitary gates and projective measurements drives a \emph{coding transition}: the volume-law phase functions as a robust quantum code, protecting logical information for timescales exponential in the system size. This robustness is dynamically signaled by a divergence in the purification time, reflecting the system's capacity to retain quantum correlations with an initially entangled reference \cite{Gullans_dynamical_purification_2020, Choi_QEC, gyawali_2025_simulating, gyawali_2026_coding_phases}. Formally, this capacity to transmit quantum information is captured by the \emph{coherent information}~\cite{schumacher_1996_dataprocessing}---the quantum analogue of the Holevo information that governs the MILT. The two transitions are thus distinguished at a fundamental level: the MILT and MIPT correspond to transitions in the classical and quantum channel capacities of the circuit, respectively.

\begin{figure}[th]
    \centering
    \includegraphics[width=0.45\textwidth]{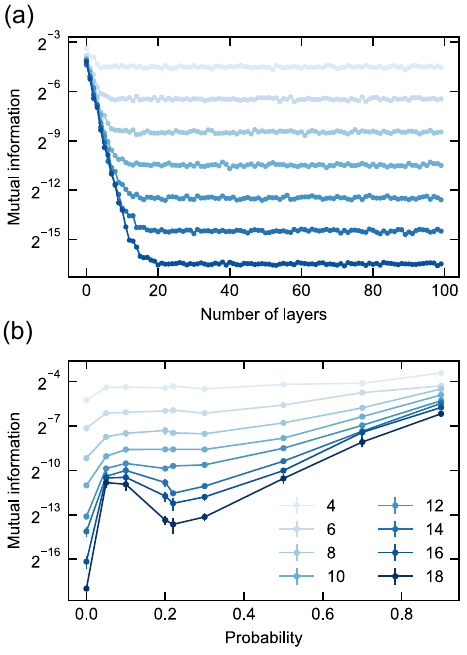}
    \caption{\textbf{Mutual information between Alice and Bob.} \textbf{a} An exponential drop of mutual information $I(A:B)$ between Alice and Bob in the HEA2 ansatz with no measurements, Alice drawing parameters ${\boldsymbol\theta}$ uniformly and Bob measuring $O = Z_0 Z_1$. These results establish the BP as a no-coding phase without measurements. \textbf{b} Mutual information for the equivalent monitored circuit at layer $N$, the number of qubits, post-selected for a specific set of intermediate measurement outcomes,  showing three distinct regions: the mutual information immediately jumps for a finite but small $p$ dips until around 0.2 and grows again. The second transition coincides with the MILT observed in the cost-function gradients.}
    \label{fig:mutual_info}
\end{figure}

We now identify the MIPT in our three ansatzes. To do so, we rely on computing the subsystem von Neumann entropy to identify the entanglement phase transition.  We note that two of them have been studied previously by Wiersema et al.  \cite{wiersema2023measurement}. The entropy is calculated for the post-selected states and averaged over various circuit realizations and measurement outcomes for the MIPT is absent if measurement outcomes are ignored. The entanglement entropy calculation was carried out for system sizes $N=6-18$. Detecting the critical point, therefore, requires finite-size scaling analysis. Inspired by prior success  \cite{Skinner_MIPT_2019, Li_MIPT_2019, Gullans_dynamical_purification_2020, Zabalo_critical_properties_2020,wiersema2023measurement, Jian_measurement_induced_criticality_2020}, we use the scaling ansatz of the following form
\begin{equation}
    \label{eqn::scaling collapse}
    S(p,N) - S(p_c^{\mathrm{MIPT}},N) = F((p-p_c^{\mathrm{MIPT}})N^{1/\nu}),
\end{equation}
where $F(x)$ is the scaling function. In \cref{fig::scaling_analysis} of \cref{app:finite_size_scaling}, we plot the left-hand side as a function of $(p-p_c^{\mathrm{MIPT}})N^{1/\nu}$ and observe that the curves for various system sizes collapse into a single curve. Notably, these collapses are consistent with a linear scaling in $p-p_c^{\mathrm{MIPT}}$. The critical parameters extracted from the numerics are listed in \cref{tab:critical_parameters}, and we provide further details in \cref{app:finite_size_scaling}. These results show that the MIPT occurs at an ansatz-specific critical value of $p$, which is up to 4.8 times larger than $\pcmiltrange$ where the BP collapses (see  \cref{sec:variance_study}).

\begin{table}

\begin{tabular*}{\columnwidth}{@{\extracolsep{\fill}} lcc @{}}
 \toprule
 \textbf{Ansatz} & $p^{\text{MIPT}}_c$ & $\nu$ \\ 
 \midrule
 HEA1 & $0.42 \pm 0.02$ & $1.18 \pm 0.03$  \\ 
 HEA2 & $0.48 \pm 0.01$ & $1.25 \pm 0.04$  \\
 XXZ-HVA & $0.27 \pm 0.02$ & $1.26 \pm 0.06$ \\
 \bottomrule
\end{tabular*}

\caption{\textbf{Critical measurement rate $p^{\text{MIPT}}_c$ and correlation length exponent $\nu$} The results were obtained numerically by finite size scaling of the von Neumann entropy shown in SI.}
\label{tab:critical_parameters}
\end{table}

We now turn to determine the information-theoretic properties of optimization, viewing it as a communication channel as discussed in \cref{sec:infotheory}. To do so, we will compute the mutual information $I(A:B)$ between Alice's classical information and Bob's measurement results following the discussion in the methods section, \cref{sec:IAB_methods}. There are many ways of defining mutual information depending on (i)how Alice encodes her information into the quantum circuit, (ii) local operations + measurements that Bob can access to recover the classical information (iii) and whether the intermediate measurements are known to Bob. The degree of information encoded in the variational parameters that can be recovered in the cost function determines the trainability of the circuit. Thus, we consider the case in which Bob can measure the operator $Z_0Z_1$ with outcomes $i=-1,1$ with a conditional probability $p(i \mid \boldsymbol{\theta})$. Similarly, Alice has access to all the parameters $\boldsymbol{\theta}$ of the HEA2 ansatz and chooses them with a uniform probability $p_A(\boldsymbol{\theta})$.
Furthermore, we can insert intermediate measurement gates to the variational circuit as depicted in \cref{fig:ansatzes} and consider measurement outcomes $\mathbf{M}$. We compute mutual information $I(A:B)$ for HEA2 circuits, with or without mid-circuit measurements, in \cref{fig:mutual_info}. When mid-circuit measurements are present, we consider the case where measurement outcomes are post-selected and unknown to Bob for the circuits with intermediate measurements. We note that the mutual information reported here is not the same as the $I_{\text{acc}}$ from \cref{eqn:I_acc} since we neither optimize over $p_A(\boldsymbol{\theta})$ nor Bob's POVMs $\{E_i \}$. This choice was made for direct comparison to the variance results where we uniformly average over the variance of a local cost function. Now, we now discuss our mutual information results in \cref{fig:mutual_info}, beginning with the unmonitored case.

\cref{fig:mutual_info}(a) shows that in the absence of measurements, $I(A:B)$ exhibits an exponential decay with system size for circuits of depth $\gtrsim N$, mirroring the behavior of the cost gradient variance. Notably, the mutual information figure is nearly identical to the original results establishing the existence of BPs \cite{McClean_barren_plateaus_2018}, with one quantitative difference: $I(A:B)$ saturates at a circuit depth approximately $1/10$ of that required for the gradient variance to saturate, making it a more sensitive diagnostic of the BP. These results confirm that the mutual information captures the onset of the barren plateau in the same way as the gradient variance.

\begin{figure}[ht]
    \centering
    \includegraphics[width=0.45\textwidth]{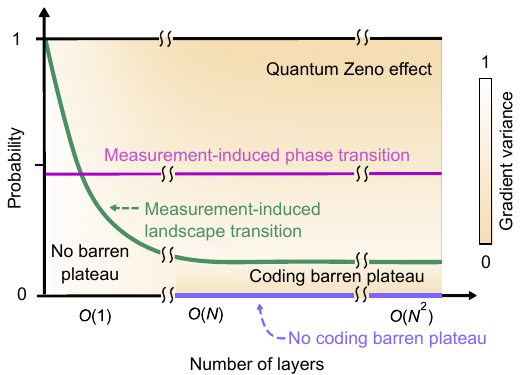}
    \caption{\textbf{Phase diagram of monitored variational quantum circuits.} The variance of the gradient shows the measurement-induced landscape transition (MILT) (shown with a dark green line) between the BP and the quantum Zeno regions. The variance of the gradients is maximum in the crossover region, and optimization here is the easiest. The dotted magenta line shows the measurement-induced phase transition (MIPT) for the HEA2 ansatz. Mutual information, a classical information-theoretic quantity connecting parameters to measurement outcomes, however, shows a BP only near $p=0$ (shown in solid purple line), which we call a no-coding BP as Bob cannot recover the classical information encoded by Alice.}
    \label{fig:phase_diagram}
\end{figure}

\cref{fig:mutual_info}(b) illustrates the mutual information in the presence of measurements, post-selected on the specific outcome set $\mathbf{M} = \{0, 0, \ldots, 0\}$. The choice of $\mathbf{M}$ does not affect the results, since we average over random circuit parameters. We identify three distinct regions separated by two transitions: one at $p = 0$ and a second at $p \approx 0.2$. At low but finite measurement rates $0 < p \lesssim 0.2$, the mutual information rises sharply and becomes independent of system size, despite the gradient variance remaining exponentially suppressed. This regime constitutes a \emph{coding barren plateau}: although optimization is obstructed, a code exists, though possibly difficult to identify, through which the cost function retains sensitivity to changes in the parameters. Near the second transition, the system-size dependence re-emerges as a dip in $I(A:B)$, and beyond it,  the mutual information survives but its overall scale decreases linearly with measurement probability. The second transition aligns closely with the MILT identified in the gradient variance and optimization traces, corroborating its identification as a distinct critical point. Thus, the mutual information reveals that trainability in the presence of a BP can be improved at finite but low measurement rates under post-selection, and sharpens the distinction between coding and non-coding barren plateau phases in the phase diagram of \cref{fig:phase_diagram}. If a manageable trainability can be achieved at a logarithmic-in-system size measurement rate, then the VQA would achieve a polynmomial instead of exponential post-selection overhead.

\subsection{Statistical mechanics mapping and the stability of the BP}
\label{subsec:stat_mech_mapping}
To obtain analytical insight to the above numerical results, consider now our fourth ansatz, the HUA. Here each layer is composed of arbitrary 2-qubit unitary gates randomly chosen from the Haar measure and arranged in a brickwork fashion, with measurements on each qubit after any layer with probability $p$. In the absence of measurements, the average gradient $\langle \partial C\rangle$ vanishes and the variance of the gradient is quadratic in the unitary circuit 
\begin{equation}
    \text{Var}(\partial C) = \int dU \left(\langle0|(\partial U^\dagger H U+U^\dagger H\partial U\right)|0\rangle^2.
\end{equation}
Hence, it can be computed using the Weingarten integrals \cite{collins2003moments,collins2006integration, dankert_2009_exact_and_approximate, mele_2024_intro_to_haar_measure_tools} and enables analytic insight into the barren plateau problem much like how the HUA was shown to produce $t$-designs for sufficiently long circuits \cite{hunter2019unitary}. As discussed in \cref{app:state_mech_model}, we find the barren plateau involves $\text{Var}(\partial C)$ which is different from the frame potential that governs the $t$-design case, though both involve an Ising-like state mech model with domain walls. The frame-potential has periodic boundary conditions, a uniform bulk, and a single fluctuating domain wall. On the other hand, $\text{Var}(\partial C)$ has a free initial time boundary, a fixed final time boundary, and two special locations, the gradient and a local term in the cost function. It also governed by the generation of two domain walls in the bulk that terminate at the cost function on the final time boundary. Hence, though the $t$-design property of the circuit and the growth of entanglement is important for the establishment of a BP, we cannot directly import results from the frame potential to understand them. 

In the presence of post-selected measurements, the cost function becomes a fraction $C = N/D$, where the numerator $N$ is the expectation value in the collapsed wave function produced by the measurements and the denominator $D$ is needed to normalize the state. Thus $C$ is no longer quadratic in the unitary gates $U$. To evaluate this cost function and its gradients, we can use two approaches. One is the replica trick, which is a general solution, and the other is to work in the limit of a dilute set of measurements. We choose the latter, a rigorous explanation of which is presented in \cref{subsubsec:the_role_of_measurements}. In this limit, $D = q^{-m} + R$, where $q=2$ is the dimension of a single qubit (or qudit for arbitrary $q$), $m$ is the number of measurements and the residual $R$ is small. In \cref{thm:cost_function}, we prove that $\langle C\rangle$ is dominated by $\langle N\rangle/\langle D\rangle$ as $q\to\infty$ and then in \cref{conj:ave_variance}, we argue this expansion should extend to $Var(\partial C)$. Specifically, we conjecture the variance is approximated by 
\begin{equation}
  \mathrm{Var}(\partial C) = q^{2m}\langle(\partial N)^2\rangle + \epsilon
\end{equation} 
where $\epsilon$ captures the subleading terms in the expansion, the first leading terms of which are ($-2^{2m+1}\langle\partial N R\rangle-2^{3m+1}\langle N\partial N\partial R\rangle$). Hence, we can focus exclusively on $\langle (\partial N)^2\rangle$, which is again quadratic in $U$ and $U^*$ and are able to use the same Weingarten integral techniques we used above to analyze the presence of a dilute number of measurements.

\begin{figure}[t]
    \centering
    \includegraphics[width=\linewidth]{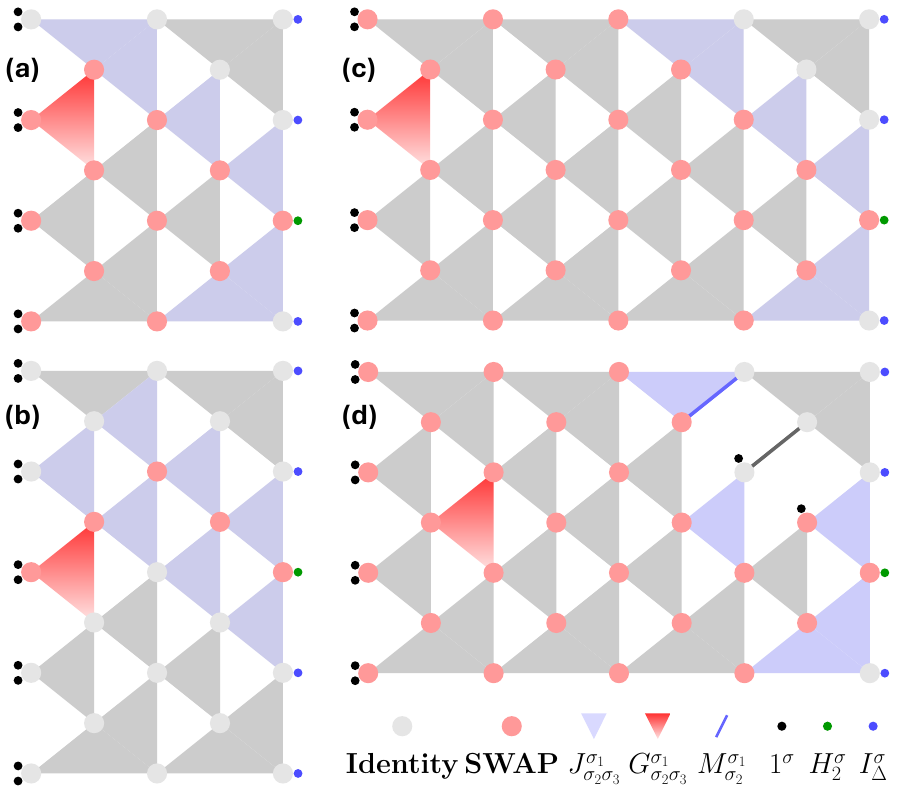}
    \caption{{\bf Mapping the HUA to a statistical mechanics model.} (a) an $N=8$ qubit circuit with $L=5$ layers of 2-qubit Haar random unitaries denoted by the gray or pink colored circles with the gradient taken on first gate of the second layer which has the effect of changing the nature of the triangle tensor, here colored red and given a gradient shading. The configuration here is the most dominant contribution for this system size for all $q\geq 2$ and shows a complex SWAP domain surrounded by the identity domain. (b) The same situation as (a) but with $N=10$. This situation now shows a clear pattern, an ``S-tube'', a width-one path of SWAP nodes surrounded entirely by identity nodes. (c) The same situation as (a) but now extended to $L=9$ layers. Here the SWAP domain fills the entire space and interface with the identity domain along straight diagonal domain walls. (d) Two measurements are placed along the diagonal domain wall, replacing triangles with a black dot and blue line with the blue line crossing the domain wall in (c) and causing it to reroute around the measurements, locally increasing the length of this domain wall but not changing where the domain wall terminates at the top border of the circuit.}
    \label{fig:statemechBP}
\end{figure}

The result of our analysis, whose details are in \cref{app:state_mech_model}, is a statistical mechanics model with negative contributions to the gradient, much like the expectation value of an observable in statistical mechanics. Following Hunter-Jones \cite{hunter2019unitary}, the model can be described by a simplical-complex diagram such as those shown in \cref{fig:statemechBP} which is the dual of more traditional tensor network diagrams. The indices of each tensor take on just two values: $I$ for ``identity'' and $S$ for ``SWAP'' which correspond to the elements of the permutation group $S_2$. This structure arises because the integral factorizes over each gate, reducing the problem to computing individual local tensors across the network and because the Weingarten integrals are quartic in $U$.

In \cref{app:state_mech_model}, we compute all of the tensors needed to describe $\text{Var}(\partial C)$, capturing them with 0-simplicies (one-index tensors), 1-simplicies (two-index tensors) and 2-simplicies (three-index tensors). The central bulk tensor is the Hunter-Jones tensor $J^{\sigma_1}_{\sigma_2\sigma_3}$ and is non-negative. The gradient $\partial$ appears as an insertion of an additional tensor appearing in the computation of $J^{\sigma_1}_{\sigma_2\sigma_3}$, producing $G^{\sigma_1}_{\sigma_2\sigma_3}$ which vanishes unless $\sigma_2=S$ in the scenario we focus on and can be negative or positive. On the boundaries we find 0-simplicies, with free boundaries at the initial time slice (either $I$ or $S$ allowed with no cost) and fixed boundary conditions at the final time (just $I$) except at the ``Hamiltonian site'' where either $I$ or $S$ are allowed with different costs and signs.  Finally, a measurement introduces a different insertion and breaks $J^{\sigma_1}_{\sigma_2\sigma_3}$ into a 0-simplex (the ones-vector $1_{\sigma_2}=1$ and a non-negative 1-simplex $M^{\sigma_1}_{\sigma_3}$.

The variance of the gradient thus maps to a statistical mechanics model with free ``initial time'' boundary conditions, open spatial boundaries, and fixed ``final time'' boundary except for sites in the support of the Hamiltonian. ``Time evolution'' then arises through the $J$ tensors, occasionally broken up by measurements $M$-tensors and once is interrupted by the gradient tensor $G$. Since the final time boundary demands the existence of $I$ sites while the gradient $G$ demands the existence of an $S$ site, both types of sites are required in a non-zero contribution to the variance of the gradient.

Figure \ref{fig:statemechBP} presents our results for the statistical mechanics model at a informal level. In the appendix, we discuss more rigorously how we obtained these results. Largely, we focus on the large $q$ limit where a single or a few configurations dominate the variance of the gradient enabling for simple conclusions.

We find this model reproduces known results for the barren plateau in variational quantum circuits. Consider the following sequence of systems. In Fig. \ref{fig:statemechBP}(a), we find by a numerical search, a complex two-domain situation dominates at large $q$. Here the number of time layers is 5 and the number of qubits 8. If we then increase the number of qubits we enter the short time shallow depth circuit case and a single ``S-tube'' extends from the Hamiltonian tensor to the gradient tensor (Fig. \ref{fig:statemechBP}(b)). If we instead increase time, i.e. the number of layers, we reach the opposite regime, where the S-domain swallows up the whole of space except for two $I$-domains emanating out of the Hamiltonian near the final time boundary to either side (Fig. \ref{fig:statemechBP}(c)). Between the $I$-domain and the $S$-domain is a \emph{domain wall}. This domain wall is stiff. It extends from the Hamiltonian term to the top or bottom of the circuit along a straight line. The domain wall cost is exponential in its length and suppresses the variance exponentially. It produces the BP for the length, as discussed in \cref{app:state_mech_model}, is proportional to $N$, the number of qubits. So, the model reproduces exponentially-in-$N$ suppression of the variance of the gradient and the absence of the barren plateau for shallow circuits \cite{McClean_VQA_2016, Cerezo_cost_function_bp_2021}.

In summary, we have been able to obtain rigorous results---for the cost function directly (\cref{thm:cost_function}) and for the variance of the gradient under \cref{conj:ave_variance}---in the presence of measurements when the measurements are dilute and take place far from where the gradient triangle is situated. This is the case presented in Fig. \ref{fig:statemechBP}(d). If two measurements are placed along the domain, it avoids the measurements, bending around them, but otherwise \emph{doesn't change the length of the domain wall}. One measurement (not shown) doesn't even change the domain wall. Hence, we have found a framework for analytically assessing variational quantum circuits in the presence of measurements and these analytical results provide strong evidence that the BP is stable in the presence of dilute measurements far from the gradient triangle. For measurements close to the gradient, the situation is more complex, yet \cref{conj:near_measurements} asserts the BP persists in this regime. 

\section{Discussion and Conclusion}
\label{sec:discussion}
Our results demonstrate a surprising role for measurements in variational quantum algorithms. We find an MILT near $\pcmiltrange$ independent of whether measurements are post-selected or ignored; and that a small number of measurements can dramatically change the classical-quantum communication channel associated with a VQA if the measurements are post-selected. Our results are largely numerical but additionally supported by a statistical mechanics model---rigorously for the cost function and conjectured for the variance of the gradient---that the BP is stable to the presence of measurements, at least if those measurements are far from the location of the gradient. We close by commenting on the role of the entanglement in a VQA with measurements, how measurements and the quantum Zeno effect compare to the traditional constant-depth circuit approach, the coding BP phase as the most important future direction, and a novel ansatz design principle suggested by our information-theoretic framework.

The role of entanglement in VQA with measurements is an important problem left unsolved by our results and suggests many questions. Volume-law entanglement between hidden and visible registers has been suggested to be the source of barren plateaus \cite{Marrero_entanglement_bp_2021, Patti_entanglement_bp_2021}. Yet, below $p_c^{\text{MIPT}}$, our system exhibits volume-law entanglement, and the variance of gradients in \cref{fig:projective_gradients} nonetheless stops vanishing with system size above $\pcmiltrange$. Hence, there is no BP, yet there is volume-law entanglement in this intermediate region. What is the role of entanglement in the origin of BP in this context? What causes the collapse of BP? One possibility is that measurements in this regime break the $t$-design property of the circuit ensemble---reducing its expressibility \cite{Holmes_expressibility_2022}---even while maintaining volume-law entanglement: the two are distinct properties, and it is the former, not the latter, that may directly control gradient variances. This further suggests, similar to the entanglement regularization of Ref. \cite{Patti_entanglement_bp_2021}, which penalizes inter-register entanglement at circuit boundaries to
mitigate BPs, the potential use of mutual information for cost-function design.

We can partly answer the collapse of the BP question by ruling out two natural suspects. First, $O$ is a linear observable and is therefore insensitive to the MIPT, so the entanglement transition alone cannot account for the BP collapse above $\pcmiltrange$. Second, the mixed gradients---for which no post-selection is performed---also exhibit the collapse, ruling out the quantum Zeno effect as the primary mechanism. A different agent is therefore at play in this region: most likely one tied to classical information flow through the optimization channel, though a deeper information-theoretic study will be needed to establish this possibility.

Using a shorter circuit is an alternative to measurements to reduce expressibility and overcome the BP. Yet the two approaches are not completely unrelated. Indeed, deep in the quantum Zeno region, intuitively, it seems like measurements cut off the earlier circuit, leaving just a region near the end whose unitaries produce the output state. Nevertheless, by introducing intermediate measurements and post-selection, we find an improvement in the gradient's variance as well as the mutual information. Our work thus motivates a deeper study of the circuit's expressibility in this region, specifically comparing the two methods of controlling entanglement.

The coding BP phase, among the three regions of the phase diagram (\cref{fig:phase_diagram}), deserves the most attention in future works. The existence of the BP in both the absence and the presence of a low rate of measurements shown in the variance studies, the cost landscape visualizations, and stat mech results (under \cref{conj:ave_variance}) provide strong evidence that the BP is stable to the introduction of measurements.  However, although the mutual information drops exponentially with system size at $p=0$, it becomes finite when $0<p<p_c^{\mathrm{MILT}}$. Furthermore, we find that mutual information is more sensitive to the circuit depth compared to the variance of the gradient in the sense that it decays to 0 faster than the gradient. Thus, we find it encouraging that the mutual information does not suffer from barren plateaus at a finite measurement rate. It is important to note that post-selection of mid-circuit measurement outcomes was necessary for this finite mutual information, although the measurement rate can remain low, making this requirement potentially scalable. This phase, therefore, exhibits a notable contradiction: the gradients vanish, yet the cost function is still somehow sensitive to parameter changes. 
 To leverage this feature, perhaps a non-gradient-based method, like ADMM  \cite{taylor_training_2016}, would perform better, or an information theoretical analysis could reveal a decoder to connect parameters to the cost function.

\section*{Data Availability}
The data supporting the findings are available on Zenodo  \cite{zenodo_milt}.
\section*{Code Availability }
The code for this study is available on Zenodo 
 \cite{zenodo_milt}. Instructions for installing dependencies and running the code can be found in the README file. For further assistance, please contact the code maintainer at str36@cornell.edu. 
\section*{Acknowledgments}
We thank Sarang Gopalkrishnan, Yong-Baek Kim, Roeland Wiersema, and Juan Felipe Carrasquilla for useful discussions.

\bibliography{main.bib}

\clearpage
\onecolumngrid
\appendix
\crefalias{section}{appendix}
\setcounter{equation}{0}
\setcounter{figure}{0}
\setcounter{table}{0}
\renewcommand{\thefigure}{S\arabic{figure}}
\renewcommand{\theequation}{S\arabic{equation}}
\renewcommand{\thetable}{S\arabic{table}}

\section*{Appendix}
\startcontents[appendix]
\printcontents[appendix]{l}{1}{}
\newpage

\section{Methods}

We developed a custom statevector simulation library in Python~\cite{zenodo_milt} to simulate the monitored quantum dynamics of variational quantum circuits. Our library implements the three types of ansatzes described in \cref{sec:ansatzes}: (a)~HEA1, (b)~HEA2, and (c)~XXZ-HVA.

The circuit evolution consists of alternating layers of unitary gates and probabilistic projective measurements. In the measurement layers, each qubit is independently measured with probability $p$ in the computational basis. This is implemented via the projection operators $P_0 = \ket{0}\bra{0}$ and $P_1 = \ket{1}\bra{1}$. For a given state $\ket{\psi(\boldsymbol{\theta})}$, the probability of outcome $0$ is $p_0 = \bra{\psi(\boldsymbol{\theta})} P_0 \ket{\psi(\boldsymbol{\theta})}$; the state is then projected to $\ket{0}$ with probability $p_0$ or $\ket{1}$ with probability $1-p_0$, and subsequently renormalized.

To estimate observables, we average over both circuit realizations (random gate placements) and measurement trajectories (stochastic outcomes). The ensemble average for an observable $O$ is approximated by:
\begin{equation}
    \label{eqn:ensemble_average}
    \mathbb{E}\left[\expval*{O}_{\mathbf{M}}\right] \approx \frac{1}{N_s}\sum_{\mathcal{C}} \expval*{O}_{\mathbf{M}|\mathcal{C}},
\end{equation}
where $\mathcal{C}$ denotes a circuit realization sampled from the ansatz distribution, and $\mathbf{M}$ represents a measurement outcome bitstring sampled from the quantum state. Unless otherwise specified, we used $N_s = 10^3$ samples. Error bars were computed using statistical bootstrapping via the SciPy library~\cite{2020SciPy, efron1993introduction}.

\subsection{Computing the gradients}
\label{app:computing_gradients}
We compute gradients for both the \emph{projective} and \emph{mixed} cost functions using analytical expressions derived below. To ensure clarity, we first define the state notation for the monitored ansatz.

The circuit consists of $L$ layers, where each layer $k$ involves a unitary operation $U_k(\boldsymbol{\theta}_k)$ followed by projective measurements. Let $\mathbf{M} = (\mathbf{M}_1, \dots, \mathbf{M}_L)$ denote the collection of measurement outcomes across all layers, where $\mathbf{M}_k$ represents the outcomes at layer $k$. The corresponding projection operator for the $k$-th layer is denoted by $P_{\mathbf{M}_k}$. We define the \emph{unnormalized} state vector conditioned on this specific trajectory as the ordered product of these operations:
\begin{equation}
    \ket{\tilde{\psi}_{\mathbf{M}}(\boldsymbol{\theta})} = \left( \prod_{k=L}^1 P_{\mathbf{M}_k} U_k(\theta_k) \right) \ket{0},
\end{equation}
The probability of observing the trajectory $\mathbf{M}$ is then given by
\begin{equation}
    p_{\mathbf{M}}(\boldsymbol{\theta}) = \braket{\tilde{\psi}_{\mathbf{M}}(\boldsymbol{\theta})}{\tilde{\psi}_{\mathbf{M}}(\boldsymbol{\theta})}.
\end{equation}
Consequently, the physical, \emph{normalized} state accessible in a post-selected experiment is $\ket{\psi_{\mathbf{M}}} = \ket{\tilde{\psi}_{\mathbf{M}}} / \sqrt{p_{\mathbf{M}}}$.

The \emph{projective} cost function is defined as the expectation value of an observable $O$ on the post-selected state
\begin{align}
    \label{eqn:operator_expectation}
    C_\mathbf{M}({\boldsymbol\theta}) = \bra{\psi_\mathbf{M}(\boldsymbol{\theta})} O \ket{\psi_\mathbf{M}(\boldsymbol{\theta})} = \frac{\bra*{\Tilde{\psi}_\mathbf{M}} O \ket*{\Tilde{\psi}_\mathbf{M}}}{p_\mathbf{M}} = \frac{\expval*{\Tilde{O}}_\mathbf{M}}{p_\mathbf{M}},
\end{align}
where $\expval*{\Tilde{O}}_\mathbf{M}$ is defined as the expectation value with respect to the unnormalized wavefunction $\ket{\tilde{\psi}_{\mathbf{M}}}$. The gradient must account for the parameter dependence of the normalization factor $p_{\mathbf{M}}$:
\begin{align}
    \label{eqn::gradient_for_an_outcome}
    \partial_{j} C_\mathbf{M}({\boldsymbol\theta}) = \pdv{C_\mathbf{M}({\boldsymbol\theta})}{\theta_j} = \frac{1}{p_\mathbf{M}}\pdv{\expval*{\Tilde{O}}_\mathbf{M}}{\theta_j} - \frac{\expval*{\Tilde{O}}_\mathbf{M}}{p_\mathbf{M}^2} \pdv{p_\mathbf{M}}{\theta_j}
\end{align}
We note that
\begin{align}
    \label{eqn:gradient_first_term}
    \pdv{\expval*{ \Tilde{O}}_\mathbf{M}}{\theta_j} = \mel{\partial_{j}\Tilde{\psi}_\mathbf{M}}{O}{\Tilde{\psi}_\mathbf{M}} + \mel{\Tilde{\psi}_\mathbf{M}}{O}{\partial_{j}\Tilde{\psi}_\mathbf{M}},
\end{align}
and
\begin{align}
    \label{eqn:gradient_second_term}
    \pdv{p_\mathbf{M}}{\theta_j} = \bra*{\partial_j \Tilde{\psi}_\mathbf{M}} \ket*{\Tilde{\psi}_\mathbf{M}} + \bra*{ \Tilde{\psi}_\mathbf{M}} \ket*{\partial_j \Tilde{\psi}_\mathbf{M}}
\end{align}
Putting them back to \cref{eqn::gradient_for_an_outcome}, we get
\begin{align}
    \partial_{j} C_\mathbf{M}({\boldsymbol\theta}) &= \frac{1}{p_\mathbf{M}} \bra*{\partial_j \Tilde{\psi}_\mathbf{M}} O \ket*{\Tilde{\psi}_\mathbf{M}} - \frac{\expval*{\Tilde{O}}_\mathbf{M}}{p_\mathbf{M}^2} \bra*{\partial_j \Tilde{\psi}_\mathbf{M}} \ket*{\Tilde{\psi}_\mathbf{M}} + \: \text{c.c.}
\end{align}
Thus, the resulting analytical gradient is:
\begin{align}
\label{eqn::projective_gradient}
     \partial_j C_{\mathbf{M}}({\boldsymbol\theta})
     = 2\text{Re} \left[ \frac{\bra*{\partial_j \Tilde{\psi}_\mathbf{M}} O\ket*{\Tilde{\psi}_\mathbf{M}}}{p_\mathbf{M}} - \frac{\expval*{\Tilde{O}}_\mathbf{M}}{p_\mathbf{M}^2} \bra*{\partial_j \Tilde{\psi}_\mathbf{M}}\ket*{\Tilde{\psi}_\mathbf{M}} \right].
\end{align}
The first term captures the change in the unnormalized state, while the second term corrects for the drift in the trajectory probability itself. This quantity can also be estimated on quantum hardware using a modified parameter-shift rule~\cite{wiersema2023measurement, Mitarai}:
\begin{equation}
    \partial_j C_{\mathbf{M}}({\boldsymbol\theta}) = \frac{1}{2} \left(\expval*{O_{\mathbf{M}}(\boldsymbol{\theta^+})} \frac{p_{\mathbf{M}}({\boldsymbol\theta}^+)}{p_{\mathbf{M}}({\boldsymbol\theta})}- \expval*{O_{\mathbf{M}}(\boldsymbol{\theta^-})} \frac{p_{\mathbf{M}}({\boldsymbol\theta}^-)}{p_{\mathbf{M}}({\boldsymbol\theta})} \right),
\end{equation}
where $\boldsymbol{\theta}^\pm$ denotes the parameter vector shifted by $\pm \pi/2$ at index $j$.

For the \emph{mixed} cost function $C(\boldsymbol{\theta}) = \sum_{\mathbf{M}} p_{\mathbf{M}} C_{\mathbf{M}}(\boldsymbol{\theta})$, which averages over all trajectories, the derivatives of the normalization factors cancel out. The gradient simplifies to the sum of the unnormalized drifts:
\begin{align}
    \partial_j C({\boldsymbol\theta}) &= \sum_{\mathbf{M}} 2 \text{Re} \left[ \bra*{\partial_j \tilde \psi_\mathbf{M}({\boldsymbol\theta})} O\ket*{\tilde \psi_\mathbf{M} ({\boldsymbol\theta})} \right].
\end{align}
Because the mixed cost is linear in the density matrix $\rho = \sum_{\mathbf{M}} \ket{\tilde{\psi}_{\mathbf{M}}}\bra{\tilde{\psi}_{\mathbf{M}}}$, its gradient follows the standard parameter-shift rule:
\begin{align}
    \partial_j C({\boldsymbol\theta}) = \frac{1}{2} \left( C({\boldsymbol\theta}^+) - C({\boldsymbol\theta}^-) \right).
\end{align}

\subsubsection{Gradients using parameter shift}
\label{app:parameter_shift}
Analytical gradients can be measured on a quantum computer using the parameter shift method. Let's take a closer look at the gradient of the unnormalized expectation value from \cref{eqn:gradient_second_term}:
\begin{align}
    \pdv{\expval*{\Tilde{O}}_\mathbf{M}}{\theta_j} = \mel{\partial_{j}\Tilde{\psi}_\mathbf{M}}{O}{\Tilde{\psi}_\mathbf{M}} + \mel{\Tilde{\psi}_\mathbf{M}}{O}{\partial_{j}\Tilde{\psi}_\mathbf{M}}.
\end{align}
Noting that
\begin{align}
    \ket{\Tilde{\psi}_\mathbf{M}} = \left(\prod_{k=L}^{j+1} P_{\mathbf{M}_k} U_k(\theta_k)\right) P_{\mathbf{M}_j} U_j(\theta_j) \left(\prod_{k=j-1}^{1} P_{\mathbf{M}_k} U_k(\theta_k)\right) \ket{0}
\end{align}
and
\begin{align}
     \ket{\partial_{j}\Tilde{\psi}_\mathbf{M}} = -\frac{i}{2}\left(\prod_{k=L}^{j+1} P_{\mathbf{M}_k} U_k(\theta_k)\right) P_{\mathbf{M}_j} G_j U_j(\theta_j) \left(\prod_{k=j-1}^{1} P_{\mathbf{M}_k} U_k(\theta_k)\right) \ket{0},
\end{align}
where $U_j(\theta_j) = \exp(-\frac{i}{2}\theta_j G_j)$. Suppose $O' = P_{\mathbf{M}_j}^\dagger \left(\prod_{k=j+1}^{L} U_k^\dagger P^\dagger_{\mathbf{M}_k}\right) O \left(\prod_{k=L}^{j+1} P_{\mathbf{M}_k} U_k(\theta_k)\right) P_{\mathbf{M}_j}$, and $\ket{\psi'} = U_j(\theta_j)\left(\prod_{k=j-1}^{1} P_{\mathbf{M}_k} U_k(\theta_k)\right) \ket{0}$. Then,
\begin{align}
     \pdv{\expval*{\Tilde{O}}_\mathbf{M}}{\theta_j} &= \frac{i}{2}\left(\bra{\psi'}G_j O' \ket{\psi'} - \bra{\psi'}O' G_j \ket{\psi'} \right)  \nonumber \\
     &= \frac{i}{2}\bra{\psi'}\left[G_j, O'\right] \ket{\psi'}
\end{align}
The standard parameter-shift rule identity states that for a generator having eigenvalues $\pm 1/2$ \cite{schuld_2019_evaluating_analytical_gradients}:
\begin{align}
    \left[G_j, O'\right]=-i\left(U_j^{\dagger}\left(\frac{\pi}{2}\right) O' U_j\left(\frac{\pi}{2}\right)-U_j^{\dagger}\left(-\frac{\pi}{2}\right) O' U_j\left(-\frac{\pi}{2}\right)\right)
\end{align}
Substituting this back,
\begin{align}
    \pdv{\expval*{\Tilde{O}}_\mathbf{M}}{\theta_j} &= \frac{1}{2} \left(\mel{\partial_{j}\Tilde{\psi}_\mathbf{M}}{O}{\Tilde{\psi}_\mathbf{M}}^{+} - \mel{\partial_{j}\Tilde{\psi}_\mathbf{M}}{O}{\Tilde{\psi}_\mathbf{M}}^{-} \right) \nonumber \\
    &= \frac{1}{2} \left( \frac{\mel{\partial_{j}\Tilde{\psi}_\mathbf{M}}{O}{\Tilde{\psi}_\mathbf{M}}^{+}}{p_{\mathbf{M}}^+}p_{\mathbf{M}}^+ \right. \nonumber \left. - \frac{\mel{\partial_{j}\Tilde{\psi}_\mathbf{M}}{O}{\Tilde{\psi}_\mathbf{M}}^{-}}{p_{\mathbf{M}}^-}p_{\mathbf{M}}^- \right) \\
    &=\frac{1}{2} \left(\expval*{O}^+p_{\mathbf{M}}^+ - \expval*{O}^-p_{\mathbf{M}}^- \right),
\end{align}
where $(.)^{\pm}$ denotes quantities calculated with the parameter shifted by $\theta_j \to \theta_j \pm \frac{\pi}{2}$. Similarly, the gradient of the normalization factor (probability) is:
\begin{align}
     \pdv{p_\mathbf{M}}{\theta_j} &= \bra*{\partial_j \Tilde{\psi}_\mathbf{M}} \ket*{\Tilde{\psi}_\mathbf{M}} + \bra*{ \Tilde{\psi}_\mathbf{M}} \ket*{\partial_j \Tilde{\psi}_\mathbf{M}} \nonumber \\
     &= \frac{1}{2} \left(p_{\mathbf{M}}^+  -  p_{\mathbf{M}}^- \right)
\end{align}
Now, writing \cref{eqn::gradient_for_an_outcome} in terms of parameter-shifted quantities,
\begin{align}
     &\partial_{j} \expval*{O}_\mathbf{M} = \pdv{\expval*{O}_\mathbf{M}}{\theta_j} = \frac{1}{p_\mathbf{M}}\pdv{\expval*{\Tilde{O}}_\mathbf{M}}{\theta_j} - \frac{\expval*{\Tilde{O}}_\mathbf{M}}{p_\mathbf{M}^2} \pdv{p_\mathbf{M}}{\theta_j} \nonumber \\
     &=\frac{1}{2}\left(\expval*{O}^+ \frac{p_{\mathbf{M}}^+}{p_{\mathbf{M}}} - \expval{O}^-\frac{p_{\mathbf{M}}^-}{p_{\mathbf{M}}} \right) - \frac{1}{2}\frac{\expval*{O}}{p_{\mathbf{M}}} \left(p_{\mathbf{M}}^+ - p_{\mathbf{M}}^- \right).
\end{align}

\subsection{Optimization and landscape visualization}
To investigate trainability, we performed optimization runs using the L-BFGS-B algorithm \cite{nocedal}. We focused on the HEA2 ansatz (8 qubits, 16 layers) across various measurement probabilities $p$. For each $p$, we generated 10 independent optimization traces starting from random initializations. We minimized two distinct cost functions: \\
\begin{enumerate}
    \item A local observable: $O = Z_0Z_1$. 
    \item The $XXZ$ Hamiltonian:
\begin{equation}
    H_{XXZ} = \sum_{i=1}^{N} \left( X_{i}X_{i+1} + Y_{i}Y_{i+1} + \Delta Z_{i}Z_{i+1} \right),
\end{equation}
with periodic boundary conditions (identifying qubit $N+1$ with qubit $1$).
\end{enumerate}

Finally, to characterize the geometry of the solution space, we visualized the cost landscape around identified local minima. Following the methodology of Ref.~\cite{li2018visualizing}, we generated 2D projections of the landscape by evaluating the cost function on a grid spanned by two random orthogonal vectors centered at the converged parameters (see \cref{fig::loss_landscape}). This visualization serves as a qualitative diagnostic for the flatness or roughness of the optimization terrain.

\subsection{Computing the mutual information}
\label{sec:IAB_methods}

In our setup, the communication protocol begins with Alice sampling a parameter vector $\boldsymbol{\theta}$ from a product distribution $p_A(\boldsymbol{\theta}) = \prod_j p(\theta_j)$, where each parameter is an independent and identically distributed (i.i.d.) variable (typically uniform over $[-\pi, \pi]$). She encodes these parameters into a quantum state via the variational circuit, effectively establishing a channel $\boldsymbol{\theta} \mapsto \rho_{\boldsymbol{\theta}}$. Bob then performs POVM measurements defined by $\{E_i\}$ and obtains the probabilities of the measurement outcomes $p(i|\boldsymbol{\theta})=\Tr[\rho_{\boldsymbol{\theta}} E_i]$. Bob's task is then to decode $\boldsymbol{\theta}$ from the distribution of his measurement outcomes. This establishes a classical information channel between Alice and Bob. If intermediate measurements occur and Bob is aware of the corresponding outcomes $\mathbf{M}$, he can utilize this additional information to determine $\boldsymbol{\theta}$. In this case, the information channel is described by $p(i,\mathbf{M}|\boldsymbol{\theta})$.

As stated in \cref{subsec:mutual_info_results}, the mutual information $I(A:B)$ can be computed in various ways. Here, we define $I(A:B)$ to capture the training of a variational circuit; namely, Alice sends information via $p(\boldsymbol{\theta})$, and Bob receives information via measurements in the eigenbasis of a local cost function $Z_0 Z_1$. If intermediate measurements occur, one can allow Bob access to the outcomes to ``decode'' $\boldsymbol{\theta}$. We note that optimizing over Alice's distribution $p(\boldsymbol \theta)$ yields the classical capacity of this classical channel \cite{shannon_1948_mathematical}. This capacity differs from the classical capacity of the quantum channel $\boldsymbol{\theta} \mapsto \rho_{\boldsymbol{\theta}}$, where further optimization over the POVMs and the encoding unitary is permitted \cite{schumacher_1997_sending_classical_information, holevo2019quantum, wilde_2013_qit}.

Below, we demonstrate how to compute an efficient sample estimator for the mutual information for each case:

\subsubsection{No mid-circuit measurements}
The probabilities of Bob's measurement outcomes in the case with no mid-circuit measurements are given by $p(i | {\boldsymbol\theta}) = \Tr[\rho_{\boldsymbol\theta} E_i]$. The mutual information is then given by $I(A:B) = S(A) + S(B) - S(A,B)$ where $S_A$ is the entropy of the distribution $p_A({\boldsymbol\theta})$, $S(A,B)$ the entropy of the distribution $p(i|{\boldsymbol\theta})p_A({\boldsymbol\theta})$ and $S(B)$ the entropy of the distribution $p_B(i) = \int d{\boldsymbol\theta} p(i|{\boldsymbol\theta})p_A({\boldsymbol\theta})$.

To compute $S(A)$ and $S(A,B)$, we must carefully treat these continuous probability distributions' entropy. We do so by discretizing the parameter space ${\boldsymbol\theta}$ into $K$ parameter values $\{{\theta}_1,{\theta}_2,\ldots,{\theta}_K\}$, each occupying a volume $\Delta{\boldsymbol\theta}$ such that $\lim_{\abs{\boldsymbol{\theta}}\to\infty} \sum_i \Delta{\boldsymbol\theta} p({\boldsymbol\theta}_i) = \int d{\boldsymbol\theta} p({\boldsymbol\theta}) = 1$. If $V$ is the volume of the continuous space and $q({\boldsymbol\theta})$ the uniform distribution over it, then $\Delta{\boldsymbol\theta} = V/K$ and $q({\boldsymbol\theta}) = 1/V$. With these definitions, the entropy for finite $K$ is approximately
\begin{equation}
S(A) = \log K - \int d{\boldsymbol\theta}\, p_A({\boldsymbol\theta}) \log\frac{p_A({\boldsymbol\theta})}{q({\boldsymbol\theta})}
\end{equation}
Similarly, we also find
\begin{equation}
    S(A,B) = \log K - \sum_i\int d{\boldsymbol\theta}\, p(i\mid{\boldsymbol\theta})p_A({\boldsymbol\theta})\log \frac{p(i\mid{\boldsymbol\theta})p_A({\boldsymbol\theta})}{q({\boldsymbol\theta})}
\end{equation}
so we can safely send $K \to\infty$ in the mutual information $I(A:B)$.

Bob's entropy $S(B)$, is computed from a discrete distribution. So we can use the standard Shannon entropy. Computing $p_B(i)$ we find
\begin{align}
p_B(i) &= \int d{\boldsymbol\theta}\, p(i\mid{\boldsymbol\theta}) p_A({\boldsymbol\theta}), \\
\end{align}
Then, we can obtain $S_B = -\sum_i p_B(i) \log p_B(i)$.

Putting these results together to compute the mutual information, we find
\begin{align}
I(A:B) &= -\sum_i\int d{\boldsymbol\theta}\, p(i\mid{\boldsymbol\theta})p_A({\boldsymbol\theta})\log\frac{p_B(i)}{p(i\mid{\boldsymbol\theta})}\\
       &= -\sum_{a,b}\frac{1}{N_aN_b} \log\frac{p_B(i_{b|a})}{p(i_{b|a}\mid{\boldsymbol\theta}_a)}
\end{align}
Where in the last line we sample $p(i\mid{\boldsymbol\theta})p_A({\boldsymbol\theta})$ by sampling $p_A({\boldsymbol\theta})$ to get ${\boldsymbol\theta}_a$, then sampling $p(i\mid{\boldsymbol\theta}_a)$ to get $i_{b|a}$ and introduce the notation $a|b$ here to remind us that sample $i_b$ was obtained by first obtaining ${\boldsymbol\theta}_a$ then sampling $p(i\mid{\boldsymbol\theta}_a)$. We also sent $K \to\infty$, so the result is exact.

\subsubsection{Accessible mid-circuit measurements}

Let us now consider the case where the channel undergoes intermediate measurements, and Bob has access to these measurement outcomes. Here, the information channel is $p(i,\mathbf{M}\mid{\boldsymbol\theta}) = p(i|\mathbf{M},{\boldsymbol\theta})p(\mathbf{M}\mid{\boldsymbol\theta})$ where $\mathbf{M}$ denotes measurement outcomes. Given such a channel, we can again compute the mutual information as in the above noiseless case. So we can directly obtain $I(A:B)$ by sending $i \to i,\mathbf{M}$. The result is 
\begin{align}
    I(A:B) &= -\sum_{i,\mathbf{M}}\int d{\boldsymbol\theta}\, p(i,\mathbf{M}\mid{\boldsymbol\theta})p_A({\boldsymbol\theta})\log\frac{p_B(i,\mathbf{M})}{p(i,\mathbf{M}\mid{\boldsymbol\theta})} \nonumber\\
       &= -\sum_{a,b,c}\frac{1}{N_aN_bN_c} \log\frac{p_B(i_{c|a,b},\mathbf{M}_{b|a})}{p(i_{c|a,b},\mathbf{M}_{b|a}\mid{\boldsymbol\theta}_a)}
       \label{eq:aware_mutual_info}
\end{align}
where the last line is the sample estimator for $I(A:B)$ given ${\boldsymbol\theta}_a$ drawn from $p_A({\boldsymbol\theta})$, $\mathbf{M}_{b|a}$ drawn from $p(\mathbf{M}\mid{\boldsymbol\theta}_a)$ and $i_{c|a,b}$ drawn from $p(i\mid \mathbf{M}_{b|a},{\boldsymbol\theta}_a)$.

\subsubsection{Fixed mid-circuit measurements}
We also consider the case where the intermediate measurement outcomes are post-selected to have a specific outcome, but Bob is unaware of the measurement outcomes. This is the case closest to a training problem because Bob is recovering all the information from the cost function and not the intermediate measurement probability distribution $p(\mathbf{M})$. The mutual information is then given by \cref{eq:aware_mutual_info}, but the summation over $b$ is replaced by a single set of measurement outcomes $\mathbf{M}$:

\begin{align}
    I(A:B) &= -\sum_{i}\int d{\boldsymbol\theta}\, p(i,\mathbf{M}\mid{\boldsymbol\theta})p_A({\boldsymbol\theta})\log\frac{p_B(i,\mathbf{M})}{p(i,\mathbf{M}\mid{\boldsymbol\theta})} \nonumber\\
       &= -\sum_{a,b}\frac{1}{N_aN_b} \log\frac{p_B(i_{c|a,b},\mathbf{M}_{b|a})}{p(i_{c|a,b},\mathbf{M}_{b|a}\mid{\boldsymbol\theta}_a)}.
       \label{eq:unaware_mutual_info}
\end{align}

\subsubsection{Marginalized mid-circuit measurements}
Here, we marginalize over the environment's measurement outcomes. The channel is then given by $p(i\mid{\boldsymbol\theta}) = \sum_\mathbf{M} p(i,\mathbf{M}\mid{\boldsymbol\theta})$. Hence, we can directly insert this into the mutual information of the noiseless case to obtain
\begin{align}
I(A:B) &= -\sum_i\int d{\boldsymbol\theta}\, p(i\mid{\boldsymbol\theta})p_A({\boldsymbol\theta})\log\frac{p_B(i)}{p(i\mid{\boldsymbol\theta})}\\
       &= -\sum_{a,b}\frac{1}{N_aN_b} \log\frac{p_B(i_{b|a})}{p(i_{b|a}\mid{\boldsymbol\theta}_a)}
\end{align}
but now we need to estimate $p_B(i_{b|a})$ and $p(i_{b|a}\mid{\boldsymbol\theta}_a)$. These are
\begin{align}
    p(i_{b|a}\mid{\boldsymbol\theta}_a) &= \sum_\mathbf{M} p(i_{b|a}\mid \mathbf{M},{\boldsymbol\theta}_a)p(\mathbf{M}\mid{\boldsymbol\theta}_a)  \\
    &= \sum_c\frac{1}{N_c} p(i_{b|a}\mid \mathbf{M}_{c|a,b},{\boldsymbol\theta}_a)\\
    p_B(i_{b|a}) &= \int d{\boldsymbol\theta}\, p(i_{b|a}\mid{\boldsymbol\theta})p_A({\boldsymbol\theta})\\
      &= \sum_{c} \frac{1}{N_c}p(i_{b|a}\mid{\boldsymbol\theta}_c).
\end{align}

\newpage
\section{Modeling Gradient Variance with Statistical Mechanics}
\label{app:state_mech_model}
\begin{figure*}[b]
    \centering
    \includegraphics[width=0.85\linewidth]{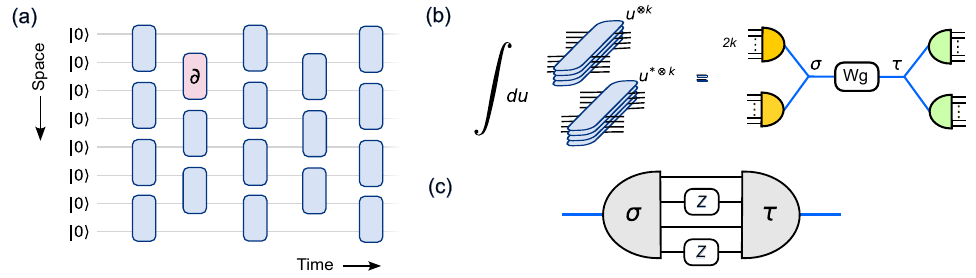}
    \caption{\textbf{The Haar random unitary brickwork model.} (a) the circuit diagram with 2-qubit unitaries, initial state, and the location of the gradient marked on one of the gates. (b) the tensor network produced by taking the Haar average over one of the gates in the superoperator construction with four external $2k$ qubit bundle leg (black) and internal permutation indices lines (blue). (c) the gradient insertion between two $2k$ qubit bundles appearing between a gate $u$ and the gradient gate $\partial u$, here shown one of the $2^k=4$ terms for the $k=2$ case relevant for the variance of the gradient computation.}
    \label{fig:Haar_averaging}
\end{figure*}

Consider the brickwork Haar model of Fig. \ref{fig:Haar_averaging}(a), consisting of two-qubit unitaries drawn uniformly from $U(4)$ according to the Haar measure, and the role of measurements in this circuit. While maintaining rigorous analytical insight, we will first consider the case without measurements and construct a statistical mechanics model associated with the variance of the gradient. Then we will add measurements and see how they change the statistical mechanics model. In the main text, we described the results from analyzing the model. 

We can describe the brickwork model symbolically with the sequence $U_t$ of unitaries in which for $t$ odd, $U_t = \prod_{i=1, i\, \text{is odd}}^{N}u_{i,i+1}(t)$ while for $t$ even we have $U_t = \prod_{i=2, i\, \text{is even}}^{N-1}u_{i,i+1}(t)$. Each of the $u_{i,i+1}(t)$ (assuming $i$ is odd if $t$ is odd or $i$ is even if $t$ is even) is a two-qubit unitary matrix drawn from the Haar measure over $U(4)$. To view them as a variational circuit, we express them using the KAK decomposition (see Ref. \cite{zhang2003geometric} Eq. 29)
\begin{equation}\label{eq:SU4parameterization}
    u_{i,i+1}(t) = W_{iL}\!(t)\otimes W_{(i+1)L}\!(t) V_{i,i+1}(t) W^\dagger_{iR}\!(t)\otimes W^\dagger_{(i+1)R}\!(t)
\end{equation}
where single qubit unitary $W_{jL/R}(t)=e^{-i\alpha_{jL/R}(t)Z}e^{-i\beta_{jL/R}(t)Y}e^{-i\gamma_{jL/R}(t)Z}$ are defined with Euler-angles and $V_{i,i+1}(t) = e^{ix_{i,i+1}(t)X\otimes X}e^{iy_{i,i+1}(t)Y\otimes Y}e^{iz_{i,i+1}(t)Z\otimes Z}$. 

The cost function is $C(\{\theta\}) = \langle 0|U^\dagger HU|0\rangle$ where $U = U_TU_{T-1}\ldots U_1$ and we denote $\{\theta\}$ to be the set of parameters indexed by $\theta$. The gradient is then $\partial C/\partial\theta$ and its average is
\begin{equation}
   \left\langle\frac{\partial C}{\partial\theta}\right\rangle = \int dU \langle 0|U^\dagger H\frac{\partial U}{\partial\theta}|0\rangle + c.c.
\end{equation}
Introducing a superoperator formalism, we can express the average as
\begin{equation}\label{eq:gradHaar}
   \left\langle\frac{\partial C}{\partial\theta}\right\rangle = \int dU \langle\langle H|\left(\frac{\partial U}{\partial\theta}\otimes U^*+U\otimes\frac{\partial U^*}{\partial\theta}\right)|0\rangle\rangle
\end{equation}
where $|0\rangle\rangle\leftrightarrow |0\rangle\langle 0|$, $|H\rangle\rangle\leftrightarrow H$, and inner product $\langle\langle0|H\rangle\rangle = \text{Tr}\big[|0\rangle\langle 0|H\big]$. 

In a similar way, we can express the variance of the gradient as a Haar average. It is given by
\begin{equation}
\text{Var}\left(\frac{\partial C}{\partial\theta}\right) = \left\langle\left(\frac{\partial C}{\partial\theta} - \left\langle\frac{\partial C}{\partial\theta}\right\rangle\right)^2\right\rangle
= \left\langle\left(\frac{\partial C}{\partial\theta}\right)^2\right\rangle,
\end{equation}
where we assume $\langle\partial C/\partial\theta\rangle = 0$ (see below). This form means we now need to compute the Haar average
\begin{equation}
  \text{Var}\left(\frac{\partial C}{\partial\theta}\right) = \int dU \Bigg(\langle\langle H|\frac{\partial U}{\partial\theta}\otimes U^*|0\rangle\rangle +
  \langle\langle H|U\otimes\frac{\partial U^*}{\partial\theta}|0\rangle\rangle\Bigg)^2.
\end{equation}
As before, we can convert the square to a superoperator space, here achieved by doubling, expressing the integral as
\begin{equation}\label{eq:varHaar}
  \text{Var}\left(\frac{\partial C}{\partial\theta}\right) =  \int dU \langle\langle H|\otimes \langle\langle H|\left(\frac{\partial U}{\partial\theta}\otimes U^*+U\otimes\frac{\partial U^*}{\partial\theta}\right)^{\otimes 2}|0\rangle\rangle\otimes |0\rangle\rangle
\end{equation}
So, we can express the Haar average gradient and variance as a time evolution in a superoperator space beginning from the all-$|0\rangle$'s state and ending at the Hamiltonian ``state''. 

\subsection{Computing the Haar Averages}
The integrals over $U$ in Eqs. \eqref{eq:gradHaar} and \eqref{eq:varHaar} breaks down into products of integrals over the individual gates $u_{i,i+1}$ and so we can focus on the operators $\int du u\otimes u^*$ for the gradient and $\int du (u\otimes u)^{\otimes 2}$ for the variance of the gradient, where we now suppress indices for notational simplicity. In each case, there is one special gate where the gradient takes place, namely where $\partial u/\partial\theta\neq 0$. Keeping to the simplest case, we choose $\theta$ such that this derivative produces the operator $-i(Z\otimes I - I\otimes Z)$, which makes $\theta =\alpha_{iL}$ or $\alpha_{iR}$, or $\gamma_{iL}$, or $\gamma_{iL}$. Hence, we arrive at
\begin{equation}
\int du (\partial u\otimes u^* + u\otimes \partial u^*) = -i(Z\otimes I-I\otimes Z)\int du u\otimes u^*.
\end{equation}
and the gradient amounts to an \emph{insertion} between gates.

To compute $\int du (u\otimes u^*)^{\otimes k}$, the $k^{\text{th}}$ moment Weingarten integral over $U(d) = U(q^2)$ where $q=2$ is the qubit case, we use\cite{collins2003moments,collins2006integration}
\begin{equation}\label{eq:weingarten_full}
    \int du u_{i_1j_1}u_{i_2j_2}\ldots u_{i_kj_k}u^*_{m_1\ell_1}u^*_{m_2\ell_2}\ldots u^*_{m_k\ell_k} =
    \sum_{\sigma,\tau\in S_k}\prod_{a=1}^k\delta_{\sigma}(i_a,m_a)\delta_{\tau}(j_a,\ell_a)\mathcal{W}g(\sigma^{-1}\tau,d).
\end{equation}
where $S_k$ is the symmetric group of k objects, $\delta_\sigma(i_a,j_a)=\delta(i_a,j_{\sigma(a)})$ with $\sigma(a)$ a permutation of $a$ and here the indices with subscripts $i_1$, $j_1$, etc., do not denote qubits but instead the combined input and output index of the unitary. For large $d$, arbitrary $k$ the Weingarten function is \cite{collins2003moments,collins2022weingarten}
\begin{equation}
    \mathrm{Wg}(\sigma^{-1}\tau,d) = \frac{(-1)^{|\sigma^{-1}\tau|}}{d^{k+|\sigma^{-1}\tau|}}
\end{equation}
where $|\sigma|$ is the Cayley distance, the minimal number of transpositions, swaps of two of the objects in $S_k$, needed to produced the element $\sigma\in S_k$. This leading form, computing the variance of the gradient at large $d$ or large $q$, simplifies the analysis of the problem, and is especially important at arbitrary $k$ where the Weingarten function is otherwise more complex to determine. 

For our $k=1$ moment integral, we obtain
\begin{equation}
    \int du u_{i_1j_1}\otimes u^*_{m_1\ell_1} = \delta(i_1,m_1)\delta(j_1,\ell_1) \mathrm{Wg}(I,d)
\end{equation}
where $I$ is the identity and the only element in $S_1$ and indices range over the four basis vectors $|00\rangle,|01\rangle,|10\rangle,|11\rangle$ of the Hilbert space the $u$ acts on. We can split up the delta functions to correspond to each qubit $\delta(i_1,m_1) = \delta(wx,yz) = \delta(w,y)\delta(x,z)$. In this way, we can describe the integral over k-copies of a 2-qubit unitary with a tensor network as shown in Fig. \ref{fig:Haar_averaging}(b). Here the blue lines represent permutation indices that run over the group $S_k$, while the $2k$ black bundle of lines represents the superoperator qubit indices. The semicircles are the superstates $|\sigma\rangle\rangle^k$, which are the tensors associated with the delta function $\delta(w,y)$ generalized to $k>1$ (see below for $k=2$), and the dot is a three-index Kronecker delta function $\delta_{\sigma\sigma'\sigma''}$ needed to express the sum over $\sigma$ in index notation or tensor network diagram notation (repeated indices summed over). Finally, the $\text{Wg}$-tensor represents the Weingarten function $Wg(\sigma^{-1}\tau,d)$. Hence, Fig. \ref{fig:Haar_averaging}(b) represents a tensor network that is the result of the Haar average over one gate in the variance of the gradient. 

To complete the computation of the gradient, we must consider the insertion of the $Z$-operators into one of the qubit bundles (the $k=1$ case, see Fig. \ref{fig:Haar_averaging}(c) for diagram of the $k=2$ case). This is our first example of an \emph{insertion}. The resulting qubit bundle lines associated with the contraction of the $\delta(w,y)$ single qubit tensors can be expressed as
\begin{equation}
    -i\langle\langle I|\left(Z\otimes I -I\otimes Z \right)|I\rangle\rangle = -i(\text{Tr}Z-\text{Tr}Z)=0.
\end{equation}
since for $k=1$ there is only one possible value for the permutation index and so $\delta(w,y)\to |I\rangle\rangle$ in our superoperator formalism. Hence, the average gradient vanishes as expected. 

Consider now the variance of the gradient, the $k=2$ case. The Haar integral again factorizes by gates, and the derivatives affect one gate. For the gates unaffected by the gradient, we need to compute
\begin{equation}
   I_1 = \int du u_{i_1j_1}u^*_{m_1\ell_1}u_{i_2j_2}u^*_{m_2\ell_2}
\end{equation}
which is the second moment $k=2$ case of Eq. \eqref{eq:weingarten_full}. Hence,
\begin{equation}
 I_1 = \sum_{\sigma,\tau\in S_2}\prod_{a=1}^2\delta_{\sigma}(i_a,m_a)\delta_{\tau}(j_a,\ell_a)\mathrm{Wg}(\sigma^{-1}\tau,d).
\end{equation}
Let's look at the delta functions $\delta_\sigma(i_a,m_a)=\delta(i_a,m_{\sigma(a)})$. Again, they split at the qubit level 
\begin{align}
    \delta(i_a,m_{\sigma(a)}) &= \delta(w_ax_a,y_{\sigma(a)}z_{\sigma(a)})= \delta(w_a,y_{\sigma(a)})\delta(x_a,z_{\sigma(a)}).
\end{align}
The action of the permutation $\sigma$ occurs in our augmented superoperator space, and it acts uniformly on each qubit involved in the two-qubit gate. We can represent the integral with the tensor product denoting individual qubits
\begin{equation}
    I_1 = \sum_{\sigma,\tau\in S_2} |\sigma\rangle\rangle\rangle\rangle\otimes|\sigma\rangle\rangle\rangle\rangle \: \mathrm{Wg}(\sigma^{-1}\tau,q^2) \: \langle\langle\langle\langle\tau|\otimes\langle\langle\langle\langle\tau|.
\end{equation}
This form emphasizes two tensors, the superkets/bras $|\sigma\rangle\rangle^{\otimes 2}\equiv|\sigma\rangle\rangle\rangle\rangle\leftrightarrow \prod_a\delta_{\sigma}(w_a,y_a)$ that each have $2k=4$ qubit lines and one permutation index and the Weingarten tensor $\mathcal{W}g(\sigma^{-1}\tau,q^2)$ that has two permutation indices. This gives us the tensor network diagram in Fig. \ref{fig:Haar_averaging}(b) now with permutation indices taking on two possible values which we denote $I$ for identity and $S$ for swap.

\begin{figure*}[b!]
    \centering
    \includegraphics[width=0.9\linewidth]{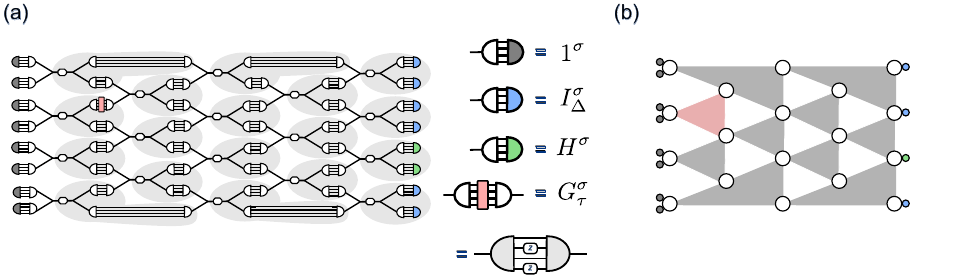}
    \caption{\textbf{Reducing the tensor network diagram into a statistical mechanics model} (a) The tensor network diagram corresponding to the average variance of the cost function for a brickwork circuit shown in \cref{fig:Haar_averaging}(a). In the bulk of the circuit, each four-legged tensor comes from integrating $2k$ copies of the brickwork circuit for $k=2$. The dark-gray semicircles at the start corresponds to the initial state $\superket{0}^{\otimes k}$. At the end, the blue semicircle corresponds to the identity operator $\superket{I}^{\otimes k}$, whereas the green semicircle corresponds to one of the Hamiltonian $Z$ operators in $k=2$ copies, i.e. $\superket{Z}\otimes\superket{Z}$. The gradient with respect to a gate parameter results in an insertion of single-qubit $Z$ operators that gets sandwiched between the two semicircles. Contracting over the set of two semicircles and a Weingarten function $Wg(\sigma^{-1}\tau,d)$, results in a single three-index Hunter-Jones triangle tensor shown in light gray. (b) The resulting simplex diagram of the network after contraction.}
    \label{fig:variance_tensor_network}
\end{figure*}

We are now ready to contract over all the qubit indicies and reduce the problem to a network involving sums just over the permutation indicies, i.e. arrive at a statistical mechanics model. We will follow Hunter-Jones and additionally contract over every other permutation index, contract over all qubit indices and permutation indices inside the gray bubles in Fig. \ref{fig:variance_tensor_network}(a) \cite{hunter2019unitary}. We will call these bubbles ``triangle tensors'' for they have three permutation indicies and we will associate them with a 2-simplex, a triangle. Every such triangle tensor is a combination of the Weingarten function $\mathrm{Wg}(\sigma,q^2)$ and two ``legs.'' The Weingarten functions for $S_2$ is $\mathrm{Wg}(\sigma,q^2) =(q^4-1)^{-1}\delta_{|\sigma|,0} - q^{-2}(q^4-1)^{-1}\delta_{|\sigma|,1|}$. The most common leg is $\langle\langle\langle\langle\tau|\sigma\rangle\rangle\rangle\rangle = q^2\delta_{|\sigma^{-1}\tau|,0}+q\delta_{|\sigma_{-1}\tau|,1}$, common because it has no insertions. The basic Hunter-Jones's triangle tensor is then
\begin{equation}
    J^{\sigma_1}_{\sigma_2\sigma_3} =\sum_{\tau\in S_2} \mathrm{Wg}(\sigma_1^{-1}\tau,q^2)
    \langle\langle\langle\langle\sigma_2|\tau\rangle\rangle\rangle\rangle
    \langle\langle\langle\langle\sigma_3|\tau\rangle\rangle\rangle\rangle
\end{equation}
See Fig. \ref{fig:variance_tensor_network} for details on how this triangle is constructed. We find exactly $J^{I}_{II} = J^S_{SS} = 1$, $J^I_{IS} = J^I_{SI} = J^S_{SI}=J^S_{IS} = q/(q^2+1)$. The elements $J^I_{SS}$ and $J^S_{II}$ vanish \cite{hunter2019unitary}. So, we have no cost for triangles that have uniform permutation indices while traingles associated with a "domain wall", where the permutation index switches from $\sigma_2$ to $\sigma_3$, have a reduced contribution of $q/(q^2+1)$.

To understand the statistical mechanics picture of the variance of the gradient, we additionally need a) the gradient qubit bundle insertion b) a local Hamiltonian ``final state'' tensor c) the identity ``final state'' tensor and d) the ``initial state'' all zeros boundary tensor. We begin by tracing out the qubit indices. The gradient bundle insertion is 
\begin{equation}
    G^\sigma_\tau = \langle\langle\langle\langle \tau|(-i)^2(Z\otimes I-I\otimes Z)^{\otimes 2}|\sigma\rangle\rangle\rangle\rangle
\end{equation} where we generalize $Z$ from $\text{diag}(1,-1)$ to arbitrary $q$ via $Z=\text{diag}(1,-1,0,0,...)$. Here we define upper indices for incoming lines and lower indices for outgoing lines. Then we find $G^I_I=G^S_I=G^I_S=0$ and $G^S_S=4q$ or using the triple delta tensor, $G^\sigma_\tau = 4q\delta_{\sigma\tau S}$. The Hamiltonian boundary tensor is 
\begin{equation}
    H^\sigma = \langle\langle\langle\langle \sigma|Z\rangle\rangle\otimes|Z\rangle\rangle.
\end{equation}
where $|Z\rangle\rangle$ is the superstate produced by the operator $Z$. This is a single qubit tensor associated with $|H\rangle\rangle\otimes|H\rangle\rangle$ and we find $H^I=0$, $H^S=2$. For $H = Z\otimes Z\otimes I^{N-2}$, two qubit bundles have this insertion on the boundary, the two green semicircles in fig. \ref{fig:variance_tensor_network}(a), the rest have the ``identity'' tensor 
\begin{equation}
    I^\sigma = \langle\langle\langle\langle\sigma|I\rangle\rangle\otimes|I\rangle\rangle
\end{equation} 
with $I^I = q^2$ and $I^S = q$. Finally, the initial all-zeros tensor $1_\sigma = \langle 0|^{\otimes 4}|\sigma\rangle\rangle\rangle = 1$ for $\sigma=I,S$. 

The tensor network that results from contraction over all qubit indices represents a honeycomb lattice (see Hunter-Jones\cite{hunter2019unitary}). Contracting over the ``$B$-sites'' of this lattice, contracting over the permutation index inside the gray bubble in Fig. \ref{fig:variance_tensor_network}(a), produces the simpler traingular lattice simplex model. The dominant triangle tensor here is Hunter-Jones' $J^{\sigma_1}_{\sigma_2\sigma_3}$ tensor, derived above. To derive the remaining gray bubbles, we need to further contract over the middle permutation index. We now turn to this task to complete the construction of the statistical mechanics model. 

A key new triangle tensor here is the gradient triangle tensor:
\begin{equation}
    G^{\sigma_1}_{\sigma_2\sigma_3} = \sum_{\tau\in S_2}\mathrm{Wg}(\sigma_1^{-1}\tau,q^2)G^\tau_{\sigma_2}\langle\langle\langle\langle\tau|\sigma_3\rangle\rangle\rangle\rangle
\end{equation}
where we abandoned Einstein summation convention in favor of having just one $\tau$ index and no triple identity tensors like $\delta^{\tau_1}_{\tau_2\tau_3}$. We find $\sigma_2=S$ for all non-zero values of $G^{\sigma_1}_{\sigma_2\sigma_3}$ and call this the \emph{fixed $S$-site of $G$}. The specific values are $G^{S}_{SS} = (4q^3/(q^4-1)$, $G^S_{SI} = 4q^2/(q^4-1)$ (a domain wall cost out of the S domain), $G^I_{SS} = -4q/(q^4-1)$ (an $S$ domain creation in the $I$ domain), and finally, $G^I_{SI}=-4/(q^4-1)$ (a domain wall cost out of the $I$ domain). So the gradient term demands at least one vertex is in an $S$ domain, the fixed $S$-site of $G$. 

On the initial and final time boundaries, we can define a one index tensors which we can depict with 0-simplicies in our diagrams. There are the ``Hamiltonian'' tensors and an ``identity tensor'' $I^{\sigma_1}$ on the final time boundary and a ones tensor on the initial time boundary. The Hamiltonian boundary tensors come in two types, depending on whether there are $|Z\rangle\rangle\otimes|Z\rangle\rangle$ states on both qubits or just one. The one qubit case is given by
\begin{equation}
    H^{\sigma}_1 = \sum_\tau \mathrm{Wg}({\sigma^{-1}\tau,q^2})H^\tau I^\tau
\end{equation}
with $H_1^I = -2q^{-1}(q^4-1)^{-1}$, $H_1^S=2q/(q^{-4}-1)$. The two qubit case, replacing $I^\tau$ with $H^\tau$, gives $H_2^I = -4q^{-2}(q^4-1)^{-1}$ and $H_2^S = 4(q^{4}-1)^{-1}$. The identity tensor, were we instead replace $H^{\tau}$ with $I^{\tau}$, is exactly $I^I = 1$ and $I^S=0$. Finally, the freedom of the initial conditions can be captured by a single index ``ones'' tensor $1^\sigma = 1$. 

Putting these tensors all together, we arrive at a simplical-complex diagram, such as that shown in fig. \ref{fig:variance_tensor_network}(b). Unlike the $t$-design case studied by Hunter-Jones, here the boundary conditions are free at the initial time boundary, gray circles or $1_\sigma$ tensors, and fixed at the final time boundary, blue circles demanding identity $I$ configurations on the boundary and green circles accepting either identity or swap $S$ configurations. In addition there is a gradient triangle in the bulk which demands the ``fixed $S$-site of $G$''. Thus, the situation is similar to but different from the $t$-design case which had periodic-in-time boundary conditions and no special tensors in the bulk (or boundary). 

\subsection{The emergence of a BP}
We can understand how a BP arises now by following Hunter-Jones and first identifying the classes of non-zero configurations and then computing the variance of the gradient in the large-$q$ limit as a sum over these configurations, with especially simple answers arising when only one configuration contributes at leading order in $1/q$. Three possible situations then arise, either one of the non-zero configurations dominates the sum, a degeneracy of such configurations dominates, or a cancelation arises between large positive weighted configurations and large negative weighted configurations, produced by either the gradient triangle or the Hamiltonian tensor, leaving subdominant configurations to determine the outcome.

\begin{definition}[Simplicial complex diagrams]
We can graphically depict our tensor expressions as a \emph{simplicial complex}. Using triangles for three-index tensors (2-simplices), lines for two-index tensors (1-simplices), and dots for 1-index tensors (0-simplices), we can glue everything together and make a ``simplicial complex'' diagram, which, if formalized, mathematically represents the variance of the gradient. We present examples of these diagrams in Fig. \ref{fig:statemechBP}. 
\end{definition}

\begin{definition}[Triangular lattice coordinates]
We can define a coordinate system for the triangular lattice using the traditional Bravais lattice coordinates with each site captured by the point $r = m a_1 + na_2$. where $a_1=(0,-1)$ and $a_2 = (\sqrt{3}/2,-1/2)$. Then placing the top left corner at $r=0$, the next site down, along the spatial axis, is at $a_1$ or $m=1$, $n=0$, the one below that is at $2a_1$, etc. so that a general site is given by $r = xa_1$ for $0<x<N/2$. This is the initial time slice. To move forward in time we have to travel along a zig-zag path, with sites at the top boundar given by $a_2$, $2a_2-a_1$, $3a_2-a_1$, $4a_2-2a_1$, etc. So the $x=0$ boundary is at $-\lfloor t/2\rfloor a_1+ta_2$, $0\leq t<T$. Similarly, the bottom boundary is at $(N/2-1)a_1 -\lfloor t/2\rfloor a_1+ta_2$. Hence, a generic point is given by $r(x,t) = (x-\lfloor t/2\rfloor)a_1 + ta_2 = (\sqrt{3}t/2,(\lfloor t/2\rfloor-t/2-x)$ with the boundaries defined by $t=0$, $t=T-1$, $x=0$ with $t$-even, and $x=N/2-1$ with $t$-even. All sites with $t$-odd are inside the bulk and not on the boundary.  
\end{definition}

\begin{definition}[$S$-domain or $I$ domain] A single contiguous domain of $S$-sites surrounded by $I$ sites, i.e. an $I$-domain, possibly touching boundaries or vice-versa. If the $S$-domain extends from one point to a point later in time, we may refer to this domain as an $S$-channel when it is more than one site thick and an $S$-tube if it is one site thick. 
\end{definition}

\begin{definition}[Domain walls and spacetime domain walls] A domain wall is any pair of neighboring nodes on the same time slice in which one node is in the $I$ configuration and the other in an $S$ configuration. A space-time domain wall is the spacetime boundary between an $S$-domain and an $I$-domain.
\end{definition}

\begin{proposition}[Exclusivity of the $S$-domain] The only class of non-zero configurations are the single $S$-domain configurations. It must include the fixed-$S$ site of the gradient triangle and the Hamiltonian site at $t=T-1$.  
\end{proposition}
\begin{proof}
Let $X=\{I,S\}^N_{nodes}$ denote the set of all assignments of $I$ and $S$ to each node. For $N$ and $T$ even, $N_{nodes} = (N-1)T/2$. The contribution of an assignment is the product of all tensor in its diagram. If one tensor is zero, the contribution vanishes. Our goal is to identify which contributions survive. 

The $S$-channels have a highly constrained behavior. If for $t<t_i$, all sites are in an $I$-domain, an $S$ site can always be inserted at a time $t$ via $J^I_{IS}$ or $J^I_{SI}$ provided the surrounding sites, if present, are in an $I$ state. This origination can occur either on the boundary or in the bulk. 

Consider the case where they only originate in the bulk. Since $J^S_{II} = 0$ and $G^S_{II} = 0$, these $S$ sites propagate in time, forming $S$-channels or $S$-nodes and cannot terminate until the Hamiltonian node at the final time boundary. Since the Hamiltonian node is a single site, the $S$-channels must merge before terminating. But $S$-channels cannot merge using the $J^{\sigma_1}_{\sigma_2\sigma_3}$ tensor.  At time $t$, two $S$-sites separated by an $I$, i.e. a sequence $(x_{i}-1,x_i, x_{i}+1)=(S,I,S)$, cannot propagate into $(x_i-2, x_i-1,x_i,x_i+1) = (I,S,S,I)$ at time $t+1$ for this would require $J^I_{SS}$ which is zero. There is only one case here and hence mergers cannot happen. Thus, $S$-channels 'scatter' off of each other or travel in parallel keeping at least one $I$ site between them. The one exception is at the gradient triangle $G^{\sigma_1}_{\sigma_2\sigma_3}$. Here $G^I_{SS}<0$ so, following the same argument but now using $G^I_{SS}$ in place of $J^I_{SS}$ a single merger between two $S$-domains is possible. Hence, if $t_g$ is time slice of the $\sigma_1$ node of the $G^{\sigma_1}_{\sigma_2\sigma_3}$ tensor, with $t_g+1$ the time slice of the $\sigma_2$ and $\sigma_3$ nodes, then at most two $S$-domains are possible for $t\leq t_g$ and these merge at $t_g+1$. Since the two $S$-domains merge into one, they are just one $S$-domain, as the proposition demands. 

The $S$-domains originating in the bulk cannot reach the spatial boundaries. At $x=0$, $t$-even, in the absence of the gradient tensor, we would need the sequence $\big(r(0,t-1), r(0,t),r(0,t+1)\big)=\big(I,S,S)$ at $t$-odd. But this demands the tensor $J^I_{SS}$ which is zero and the same tensor we needed for mergers. Hence if the gradient tensor $G^{\sigma_1}_{\sigma_2\sigma_3}$ exists on the $x=0$ boundary, then it can enable an $S$-domain to reach the boundary but otherwise, $S$-domains can at best reach $r(0,t)$ at $t$-odd with $r(0,t+1) = I$. A similar argument applies at the $x=N/2-1$, $t$-even boundary.

If an $S$ site originates at the boundary point $r(0,t)$ for some even $t$, then it will stay there propagating along via the sequence $\big(r(0,t), r(0,t+1), r(0,t+2),\dots\big)$. If it leaves the boundary, then it cannot return to the boundary by our previous argument except if the gradient tensor resides on the boundary. A similar argument at $r(N/2-1,t)$. 

Hence, we conclude there is exactly one space-time $S$-domain, demanded by the fixed $S$ node of the gradient tensor, and terminates at the Hamiltonian site, possibly existing as two separate $S$-channels for $t\leq t_g$ but merging at $t=t_g+1$, there $t_g$ is the time slice of the gradient tensors $\sigma_1$ node. 
\end{proof}

\begin{figure*}[t]
    \centering
    \includegraphics[width=\linewidth]{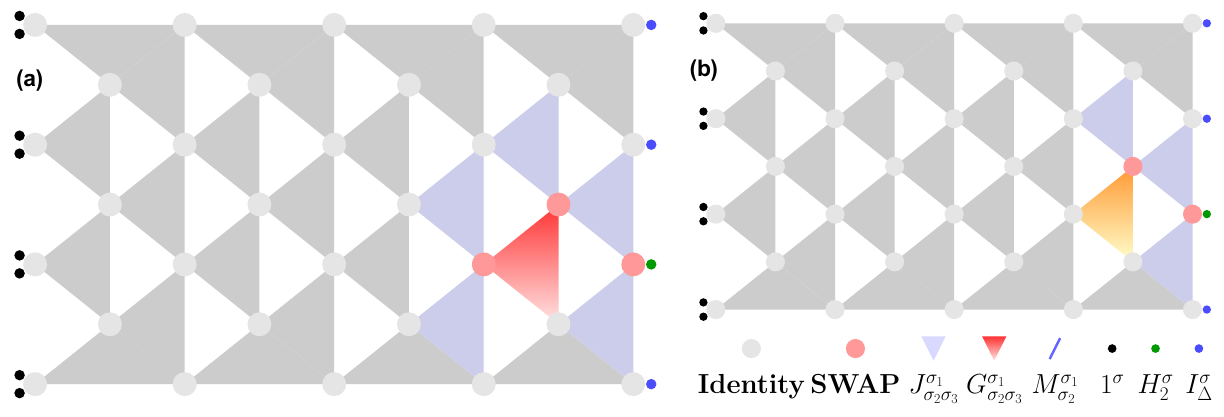}
    \caption{Here we need two diagrams at $T\gg N$ but with a gradient at late times. {\bf a} Here we see a positive contribution to the variance of the gradient with an $S$-tube forming instead of the V-shaped domain wall structure arising when the gradient is at early times. }
    \label{fig:statmech_late_times}
\end{figure*}

Now, given the highly constrained set of possible configurations, we can identify \emph{dominant configurations} at large-$q$. Fig. \ref{fig:statemechBP}(a)-(c) of the main text illustrates an evolution from a complex situation, Fig. \ref{fig:statemechBP}(a) to two simplifying limits, one where $N\gg T$, captured by Fig. \ref{fig:statemechBP}(b), and one where $T\gg N$ and the layer with the gradient happens at $t_g <T-N/2$, captured by Fig. \ref{fig:statemechBP}(c). A third limiting case is when the gradient happens at late times, $t_g > T - N/2$ (see Fig. \ref{fig:statmech_late_times}). 

\begin{definition}[Lightcone of the gradient]
    All sites within the two diagonals defined by the gradient triangle $G^{\sigma_1}_{\sigma_2\sigma_3}$, i.e. the lines captured by $\sigma_1\to\sigma_2$ and $\sigma_1\to\sigma_3$, are within the lightcone of the gradient triangle. 
\end{definition}
\begin{theorem}[No BP for $N\gg T$.]
In the $N\to\infty$, $T$-fixed limit taken such that the gradient triangle and Hamiltonian sites stay fixed as spatial boundaries are taken to infinity, and at least one Hamiltonian site exist within the lightcone of the gradient triangle, there is no BP. If the Hamiltonian is entirely outside the lightcone of the gradient triangle, the variance of the gradient vanishes.  
\end{theorem}
\begin{proof}
Assuming the Hamiltonian is within the lightcone of the gradient tensor, and we take limit defined in the theorem statement, the dominant $S$-domain configurations cannot extend to the spatial boundaries for then the domain walls between the $S$ domain and $I$ domains would be unable to converge to the Hamiltonian tensor at $t=T-1$. Hence, the only non-zero configurations are $S$-tubes or $S$-channels  existing entirely within an $I$ domain that extends to both spatial boundaries. Here, the dominant configurations are $S$-domains originating at the gradient triangle and fluctuating and/or thicking as they propagate from the gradient triangle to the Hamiltonian tensor. To compute this situation we need to count the number of all such domains and their contribution at large-$q$. But notice, this is unnecessary to prove the theorem. As we make the system larger, as we increase $N$, we do not increase the number of these configurations as we only introduce $I$ nodes in this process. Hence, consistent with the known property of shallow circuits \cite{McClean_barren_plateaus_2018}, this case has no BP.
\end{proof}

\begin{definition}[Lightcone of a Hamiltonian site]
    The straight diagonal lines connecting nearest neighbor nodes on the triangular lattice emminating from the Hamiltonian site and extending backwards in time to the spatial boundary or the initial time boundary which ever comes first. 
\end{definition}
\begin{theorem}[BP for $T\gg N$]\label{thm:BPatlargeT}
In the $T\to\infty$, $N$-fixed limit, the variance of the gradient scales like $q^{-N}$ as $q\to\infty$ and exhibits a BP. 
\end{theorem}
\begin{proof}
In this case, there is just one dominant configuration. Here, we need to understand the contribution of domain walls. Inside each domain we have either $J^I_{II}=1$ or $J^S_{SS}=1$ and hence there is no cost, no reduction in the value of a contribution due to bulk domain sites. But at domain walls, we have $J^I_{SI} = J^I_{IS} = J^S_{SI} = J^S_{IS} = q/(q^2+1)\to 1/q$ at large $q$. Hence, domain walls reduce the contribution of a configuration, a reduction proportional to the circumference of the spacetime $S$-domain. The dominant configuration is then the configuration with the fewest factors of $J^I_{SI}$, $J^I_{IS}, J^S_{SI}$, or $J^S_{IS}$, i.e. the shortest spacetime domain walls. This $S$-domain completely fills the lightcone of the Hamiltonian site with $I$ domains filling all sites outside this lightcone. All other configurations introduce a longer spacetime domain wall within the lightcone for they exhibit a slower velocity, they involve deviations of the domain wall away from the boundary of the lightcone. These configuration lengthens the domain wall, reducing their contribution at large-$q$ by additional factors of $1/q$. Since the length of the spacetime domain wall grows with $N$ for $T\gg N$, the contribution decays like $q^{-N}$ and this regime exhibits a BP. 
\end{proof}

We can now easily count the number of domain walls, the number of blue triangles for the case shown in Fig. \ref{fig:statemechBP}(c) generalized to arbitrary $N$. There are $N-2$ of them and they contribute a factor of $q^{-N+2}$ to the variance of the gradient. The gradient triangle in the $S$-domain contributes $G^S_{SS} = (4q^3/(q^4-1)\to 4/q$ and the Hamiltonian tensor contributes $H^S_2 = 4/q^4$ so that in this case we arrive at a contribution of $16q^{-N-3}$ and for fixed $q$ this exponentially vanishes in $N$ with constant $c=1>0$, as expected.

\begin{proposition}[Late-time cancellations]
A gradient triangle $G^{\sigma_1}_{\sigma_2\sigma_3}$ whose $\sigma_1$ index is at a time-slice $t > T-N/2$ enables contributions each of which decay with an $N$-independent power in $q$ as $q\to\infty$, but whose sum is bounded by $O(q^{-N})$.
\end{proposition}

\begin{proof}
Consider placing the gradient triangle at $t>T-N/2$ in the $T\gg N$ regime. Now the $S$-node configuration dominates over the rotated V-shaped $S$-domain, as shown in Fig. \ref{fig:statmech_late_times}.  It would seem like there is no barrent plateau, that the average variance is a constant independent of $N$.  However, strikingly, there is a second negative contribution at this same power of $q$ that cancels this contribution. In Fig. \ref{fig:statmech_late_times}(a) we see that an $S$-tube configuration like the one we earlier saw in Fig. \ref{fig:statemechBP}(a) is again allowed, with the gradient triangle colored red due to the positivity of $G^S_{SI}$ . In Fig. \ref{fig:statmech_late_times}(b), we find a \emph{yellow} gradient triangle, signifying the negativity of $G^I_{SI}$. Computing the contribution from Fig. \ref{fig:statmech_late_times}(a) we find at large-$q$ a value of $16q^{-11}$. This is the largest contribution to the sum, though there could be multiple of them depending on the shape of the $S$-node. In Fig. \ref{fig:statmech_late_times}(b) we find $-16q^{-11}$, exactly the same contribution but with opposite sign. For each positive contribution, there is an equal and opposite negative contribution obtained by making the same conversion of an $S$ site to an $I$ site on the $\sigma_1$ node of the $G^{\sigma_1}_{\sigma_2\sigma_3}$ tensor. These configurations therefore sum to zero at large $q$ and the variance decays with a subdominant-in-$q$ power. Given McClean et. al. have rigorously proven this regime has a BP due to the massive entanglement of the states at late times \cite{McClean_barren_plateaus_2018}, these cancellations between contributions must continue to arise among the subdominant-in-$q$ configurations for the existence of a BP demands the total sum to decay at most like $q^{-N}$ at large-$N$. 
\end{proof}

\subsection{The role of measurements}
\label{subsubsec:the_role_of_measurements}
Now, in the presence of post-selected measurements, we have to start over and rederive our expressions beginning from the cost function. Due to wave function collapse, we now have to compute a fraction. In the one-measurement case, we insert a projector that breaks up our unitary circuit, sending $U(\theta)\to U_1(\theta_1)PU_2(\theta_2)$, and then normalize the resulting state. The cost function becomes
\begin{equation}\label{eq:cost_function_with_measurements}
    C(\theta) = \frac{\langle 0|U_1^\dagger(\theta_1)PU_2^\dagger(\theta_2)HU_2(\theta_2)PU_1(\theta_1)|0\rangle}{\langle 0|U_1^\dagger(\theta_1)PU_2^\dagger(\theta_2)U_2(\theta_2)PU_1(\theta_1)|0\rangle}
\end{equation}
where $\theta = \theta_1 \cup \theta_2$. Since this is a fraction, the average over $U$ is no longer a $k$-moment Weingarten integral. 

There are at least two possible solutions to converting the fraction into $k$-moment Weingarten integrals: using the replica trick and using Levy's lemma to capture the dilute measurement limit. We will focus on the latter since it is sufficient to support the numerical results in this paper and is more straightforward. 

Writing the cost function as $C(\theta) = N(\theta)/D(\theta)$, consider expanding about the average value of $D(\theta)$ via $D = \langle D\rangle + R$. Namely, consider the geometric series
\begin{equation}
    C(\theta) = \frac{N}{\langle D\rangle + R} = \frac{N}{\langle D\rangle}\sum_{r=0}^\infty (-1)^r\left(\frac{R}{\langle D\rangle}\right)^r.
\end{equation}
This series geometric converges if $|R/\langle D\rangle|<1$ or $D> 0$. But if $D=0$, if a unitary in the circuit orients the state such that it doesn't involve the $q=0$ state $|0\rangle$ in its superposition when a measurement takes place, then $R = -\langle D\rangle$, $|R/\langle D\rangle|=1$, and the geometric series diverges. However, the chances of such a divergence arising is rare. Hence, statistically, we can truncate the series via $C(\theta) = \frac{N}{\langle D\rangle} + \delta C$, and then bound the error produced in its expectation value $\langle C\rangle$. This can be done with at most $k=2$-moment Haar random integrals as discussed below. 

If the error $\langle \delta C\rangle$ is bounded on average, then we can make the following 
\begin{conjecture}[Average variance of the gradient converges.]\label{conj:ave_variance}
    $Var(\partial C)$ is asymptotically dominated by $Var(\partial N)/\langle D\rangle^2$ in the large-$q$ limit.  
\end{conjecture}
\begin{heuristic}
    The expansion in powers of $R$ of the average cost function converges when there are a fixed number of measurement as in the large number of qudit limit after we take its Haar average. This expansion has as it's leading term $N/\langle D\rangle$. The conjecture is that this same expansion holds for $Var(\partial C)$, that $Var(\partial (N/D)) = Var(\partial N/\langle D\rangle) + error$ where {\it error} vanishes in the large-q limit relative to the first term.
\end{heuristic}

To carry out this program, we need to establish exactly how rare the occurances of $D=0$ are. It would seem like this is a job for the concentration of measure captured by Levy's lemma
\begin{equation}
    P(|D - \langle D\rangle|>t) \leq 2e^{-q^2t^2/12K^2}
\end{equation}
which is the two-sided version of equation 5.22 specific to the unitary group case in Table 5.2 in Ref. \citenum{aubrun2017alice}, where $K$ is the Lipschitz constant of $D$, defined here by the tight inequality $|D(U)-D(U')|\leq K|U-U'|_{HS}$ or $K = \sup_{U}||\nabla D||_2$, where $|...|_{HS}$ is the Hilbert-Schmidt distance and $||...||_2$ the Frobenious norm. But we understand Barren plateaus arise because the overwhelming majority of the space of quantum circuits live with $D(U)$ close to its average $\langle D(U)\rangle$, but there are extremely rare circuits where this is not the case, we would expect $K$ to be order one.  Hence, we will avoid using Levy's lemma and instead utilize 
Chebyshyev's inequality
\begin{equation}
    P(|D-\langle D\rangle|\geq \epsilon) \leq \frac{\text{Var}(D)}{\epsilon^2}
\end{equation}
for any constant $\epsilon$ \cite{grinstead2006grinstead}. So we aim to capture the error in truncating the series in terms of $\langle D\rangle$, and $Var(D)$. 

Let us first determine $\langle \delta C\rangle$. $\delta C$ itself is given by $\delta C = \frac{N}{D} - \frac{N}{\langle D\rangle}$. We can write this as
\begin{equation}
    \frac{N}{D} - \frac{N}{\langle D\rangle} = - \frac{NR}{\langle D\rangle D}.
\end{equation}
then, to estimate $\langle C\rangle$ we break up the integral into two regions, the region $\Omega$ where $D>c\langle D\rangle$ for some $0<c<1$ and its complement $\Omega^c$. Then we can express the error using the following three terms
\begin{equation}
    \langle \delta C\rangle = - \int_\Omega \frac{NR}{D\langle D\rangle}dU +\int_{\Omega^c}\frac{N}{D}dU-\int_{\Omega^c}\frac{N}{\langle D\rangle}dU.
\end{equation}
where we returned to the original expression for the $\Omega^c$ integrals. We can now compute these terms given Levy's lemma. 

To compute the first error term, we begin by recognizing $D$ satisfies $D>c\langle D\rangle$ inside $\Omega$, so that we can replace $D$ with $c\langle D\rangle$, its minimum value. For $H = Z_iZ_{i+1}$ we additionally have $|N| \leq \max_\psi\langle\psi|H|\psi\rangle = 1$. Hence, we can bound our integral by
\begin{equation}
\left|- \int_\Omega \frac{NR}{D\langle D\rangle}dU\right| \leq \int_\Omega \frac{|N||R|}{|D|\langle D\rangle}dU \leq \frac{1}{c\langle D\rangle^2}\int_\Omega |R|dU\leq \frac{1}{c\langle D\rangle^2} \int |R| dU
\end{equation}
Using Cauchy-Schwarz, we can write
\begin{equation}
    \int |R|dU \leq \sqrt{\int R^2U}\sqrt{\int dU} = \sqrt{\text{Var}(D)}
\end{equation}
Hence, we obtain
\begin{equation}
\left|- \int_\Omega \frac{NR}{D\langle D\rangle}dU\right| \leq \frac{1}{c\langle D\rangle}\frac{\sqrt{\text{Var}(D)}}{\langle D\rangle}    
\end{equation}
a polynomial expression in a small parameter $\sqrt{\text{Var}(D)}/\langle D\rangle$.

The second term contains a pole where $D\to 0$. We can handle it by introducing the cumulative probability distribution $F(x)\equiv Pr(D(U)\leq x)$. Then
\begin{align}
    \int_{\Omega^c}\frac{1}{D(U)}dU &= \int_{\Omega^c}\frac{1}{D(U)}\left[\int \delta(x-D(U))dx\right]dU\\
    &=\int_0^{c\langle D\rangle}\frac{1}{x}\left[\int \delta(x-D(U)dU\right]dx\\
    &=\int_0^{c\langle D\rangle}\frac{1}{x}F'(x)dx
\end{align}
where we swapped the order of integration using Fubini-Tonelli's theorem (both integrands non-negative). From here, we can perform an integration by parts with $u = \frac{1}{x}$, $dv = F'(x)dx$ so that $du = -\frac{1}{x^2}dx$ and $v = F(x)$ to get
\begin{equation}
    \int_{\Omega^c}\frac{1}{D(U)}dU = \frac{F(c\langle D\rangle)}{c\langle D\rangle}-\lim_{x\to 0}\frac{F(x)}{x} + \int_0^{c\langle D\rangle}\frac{F(x)}{x^2}dx
\end{equation}
To complete the analysis, we need to then understand $F(x)$.

Expressing $F(x)$ using the Heaviside step function, we find $F(x) = \int\Theta\big(x-D(U)\big)dU$. To find a bound on this quantity, it is helpful to use the following lemma.
\begin{lemma}\label{lemma:reduced_DU}
If $E(U) < D(U)$ then $Pr(D(U)\leq x) \leq Pr(E(U)\leq x$.
\end{lemma}
\begin{proof}
    If $E(U) < D(U)$ for all $U$, then we must have 
    \begin{equation}
        Pr(E(U) \leq x) = Pr(D(U)\leq x) + Pr(E(U)\leq x \mid D(U)\geq x) \geq Pr(D(U)\leq x)
    \end{equation}
\end{proof}
So to estimate $F(x)$, we need an expression for $D(U)$ and then find a simpler expression $E(U)<D(U)$ for which we can evaluate $Pr(E(U)\leq x)$.

For the case of one measurement inserted into an $N$ qubit brickwork Haar random circuit at time $t=T$ for $q=2$ qudits, $D_1(U) = \langle 0|(U^\dagger_1\ldots U^\dagger_TPU_T\ldots U_1|0\rangle$ where $P = |0><0|\otimes I_{N-1}$ is a measurement on the second qubit in the register. Tracing out all qubits except for the first two qubits let's us write $D_1(U) = \text{Tr}\rho \Pi$ where $\Pi = U^\dagger(|0\rangle\langle 0|\otimes I)U$ is a random two-qubit projector with $\text{Tr}\Pi = 2$ and $\rho$ is the density matrix $\text{Tr}_{N-2} \big[U^\dagger_{T-1}\ldots U^\dagger_1|0\rangle\langle 0|U_1\ldots U_{T-1}\big]$. We can always express $\rho$ in its eigenbasis $\rho = \sum_{x\in[0,1]^2}\lambda_x|\lambda_x\rangle\langle\lambda_x|$, where $x$ is a length 2 bitstring indexing the eigenvalues. So $D_1(U) = \sum_{x\in[0,1]^2}\lambda_x\langle\lambda_x|\Pi|\lambda_x\rangle$. Finally, we can exploit the rotational symmetry of the unitary governing $\Pi$ to rotate the basis $|\lambda_i\rangle -> |x\rangle$ where $|x\rangle$ a computational basis vector. We are now ready to apply the lemma.

Given that $\langle x|\Pi|x\rangle\geq 0$ we can drop all but the term associated with the largest eigenvalue, which we will take to be $\lambda_{00}$. In addition, the minimum value of $\lambda_{00}$ would then be $1/4$ for the maximally mixed state on 2 qubits. Hence, we have the sequence
\begin{equation}
    D_1(U) = \sum_{x\in[0,1]^2}\lambda_x\langle x|\Pi|x\rangle \leq \lambda_{00}\langle00|\Pi|00\rangle \leq \frac{1}{4}\langle00|\Pi|00\rangle \equiv E_1(U).
\end{equation}
In this way, we decoupled all the $t<T$ layers from the final layer governing $\Pi$. Letting $w = \langle00|\Pi|00\rangle$ be the weight of the $x=00$ diagonal element of $\Pi$ we see the probability $P(E(U)\leq x)$ must be of the form
\begin{equation}
    \int_0^1 dw\, p(w)\Theta(x-w/4) = \int_0^1dw\,p(w)\Theta(4x-w)
\end{equation}
where $p(w)$ is the probability density function for finding a uniformly randomly chosen projector $\Pi$ with $x=00$-component $w$. 

Notice we can write $E_1(U) = \langle00|\Pi|00\rangle = |u_{0,0}|^2 + |u_{1,0}|^2$ where $u_{ij}$ are the elements of $U\in U(4)=G$. $E_1(U)$ therefore just depends on the first column of $U$ so we can split the integral over $U$ by factoring out the $U(3)=H$ subgroup via (see Ref. \citenum{helgason2022groups}, theorem 1.9):
\begin{equation}\label{eq:halgason1.9}
    \int f(g)dg = \int_{G/H}\left(\int_{H}f(gh)dh\right)d(gH)
\end{equation}
In our case, $f(gh)=f(g)$ is independent of $h$. To make this concrete, letting $g = \begin{pmatrix}\vec{u}_0 & \vec{u_1} & \vec{u}_2 & \vec{u}_3\end{pmatrix}$ be the matrix with column vectors $\vec{u}_i$ and $h$ the matrix
\begin{equation}
    h = \begin{pmatrix}
        1 & 0 & 0 & 0\\
        0 & h_{00} & h_{01} & h_{02}\\
        0 & h_{10} & h_{11} & h_{12}\\
        0 & h_{20} & h_{21} & h_{22}
    \end{pmatrix}
\end{equation}
Our function $f$ is $\Theta\big(4x-E_1(gh)\big)$ with $E_1(g)$ just depending on $\vec{u}_0$ and not $\vec{u}_i$, $i\in [1,2,3]$. Hence, the integral over $h$ is independent of $f(gh)$ and integrates to 1. The remaining integral over $gH$ are the elements in $g$ independent of the rotations defined by $H$, namely, the whole column $\vec{u}_0$. Since the vector $\vec{u}_0$ here is normalized, we are left with
\begin{equation}
    F(x) = \int_{S^7} du_0 \Theta\big(4x-|u_{00}|^2-|u_{10}|^2\big)
\end{equation}
since the four complex numbers $\vec{u}_0$ live on $S^7$ due to the normalization and expressing the complex numbers using two real numbers. Then, breaking up $u_0$ into two $S^3$ spheres, one given by $r_1n_1$ with $n_1\in S^3$ associated with the first two complex components of $u_0$, i.e. $r_1n_1 \leftrightarrow (u_{00},u_{10})$. Similarly for the second two components we introduce $r_2n_2 = (u_{20},u_{30})$. We then find $du_0 = r_1^3 dr_1dn_1r_2^3dr_2dn_2\delta(\sqrt{{r_1}^2+{r_2}^2}-1)$. Using the solid angle or $S^3$ is $\int_{S^3} dn_1 = 2\pi^2$, and of $S^7$ is $\pi^4/3$, we obtain
\begin{equation}
    F(x) = 12\int dr_1dr_2r_1^3r_2^3\delta(\sqrt{r_1^2+r_2^2}-1)\Theta(x-r_1^2)
\end{equation}
Integrating out $r_2$, and recognizing $w = r_1^2$ with $dw = 2r_1dr_1$, we obtain
\begin{equation}\label{eq:FxN2T1}
    F(x) = \int dw \left(6w(1-w)\right)\Theta(4x-w) = 48x^2-128x^3.
\end{equation}
where we discover $p(w) = 6w(1-w)$. Hence, we find 
\begin{equation}\label{eq:second_term}
    \int_{\Omega^c}\frac{1}{D(U)}dU = \frac{F(c\langle D\rangle)}{c\langle D\rangle} + \int_0^{c\langle D\rangle} dx \frac{F(x)}{x^2} = 96c\langle D\rangle-192c^2\langle D\rangle^2,
\end{equation}
which is dominated by the first term $96c\langle D\rangle$ for small $c$. This is actually a general bound for arbitrary number of measurements, for inserting a measurement into our general single measurement expression $D_1(U)$ above arbitrary $N$, arbitrary $T$ we necessarily have $D_m(U) \leq D_1(U)$ since $D(U)$ is the norm of the unnormalized (projected) state and inserting additional measurements, inserting additional projectors, can only reduced the norm. Hence we can again apply \cref{lemma:reduced_DU}, this time using $D_m(U) \leq D_1(U) \leq E_1(U)$ and obtain exactly the same bound. Hence \cref{eq:second_term} is a general bound on $\int_{\Omega^c}D^{-1}(U)$.

To bound the third term, we begin by simplifying the integral
\begin{equation}
    \int_{\Omega^c}\frac{N}{\langle D\rangle}dU \leq Vol(\Omega^c)\frac{\text{max}_\psi\langle \psi|H|\psi\rangle}{\langle D\rangle} = \frac{Pr(D(U)\leq c\langle D\rangle)}{\langle D\rangle}
\end{equation}
where we choose $H = Z_iZ_{i+1}$ as above so it's maximum expectation value is 1. We can compute this probability using Chebyshev's inequality cited above. To do so, we choose $\epsilon = (1-c)\langle D\rangle$ so that chebyshev's inequality applies to $Pr(|D(U)-\langle D\rangle|\geq (1-c)\langle D\rangle)$. This probability is a two-sided probability expressing the weight in the tails 
\begin{equation}
    Pr(|D(U)-\langle D\rangle|\geq (1-c)\langle D\rangle) = Pr(D(U)< c\langle D\rangle) + Pro(D(U) > (2-c)\langle D\rangle)
\end{equation}
Thus we find
\begin{equation}
    Pr(D(U) < c\langle D\rangle) \leq Pr(|D(U)-\langle D\rangle|\geq (1-c)\langle D\rangle)\leq \frac{\text{Var}(D)}{(1-c)^2\langle D\rangle^2}
\end{equation}
and so the third error term is bounded by
\begin{equation}
    \int_{\Omega^c}\frac{N}{\langle D\rangle}dU \leq \frac{\text{Var}(D)}{(1-c)^2\langle D\rangle^3}.
\end{equation}

Adding all three terms together, we obtain the bound on the total error of
\begin{equation}
    |\langle \delta C\rangle| \leq \frac{1}{c\langle D\rangle}\frac{\sqrt{\text{Var}(D)}}{\langle D\rangle}+96c\langle D\rangle+\frac{\text{Var}(D)}{(1-c)^2\langle D\rangle^3}
\end{equation}
To understand the value of each of these terms, we then need to compute $\langle D\rangle$, $\text{Var}(D)$. 

Now $\langle D\rangle$ is the simplest Haar average we can consider in the present context. It involves $k=1$ moment integrals, so all indices are in the $I$ state, no Hamiltonian terms are on the boundary, and no gradient term exist. Using $\mathrm{Wg}(I,q^2) = 1/q^2$ for $k=1$, and $\langle\langle I|I\rangle\rangle = q$, we find $J^I_{II} = \mathrm{Wg}(I,q^2)\langle\langle I|I\rangle\rangle^2 = 1$ on all triangles with no measurement insertions. In the presence of one measurement insertion, we need to compute $\langle\langle I|P\otimes I|I\rangle\rangle=\text{Tr} P=1$ for $P$ is assumed to project onto just one of the $q$ possible computational basis states of a qudit. Hence a triangle involving one measurement insertion contributes $1/q$ and a triangle with two measurement insertions contributes $1/q^2$ and thus we have simply $\langle D\rangle = q^{-m}$ for $m$ measurements.  

To compute $\langle D^2\rangle$, and hence also $\mathrm{Var}(D)$, we need to consider the same case we discussed above for $\langle (\partial N)^2\rangle$ where we considered the variance of the gradient in the absence of measurements, but with the additional simplification of there also being no gradient triangle and no Hamiltonian site at the final time boundary. Now, we need to compute a new triangle, one with a measurement projector inserted. We first focus on this new tensor and then return to computing $\langle D^2\rangle$. 

Inserting a measurement projector $P = |0\rangle\rangle\langle\langle 0|$ into the $\langle\langle\sigma|\tau\rangle\rangle$ factors in the computation of the $J$-triangle, give us $\langle\langle\sigma|P|\tau\rangle\rangle=1$ for any $\sigma$ and $\tau$. Thus, this triangle becomes:
$$
\sum_{\tau\in S_2} Wg(\sigma_1^{-1}\tau,d)\langle\langle\sigma|0\rangle\rangle\langle\langle 0|\tau\rangle\rangle\langle\langle\sigma_3|\tau\rangle\rangle
$$
which after replacing the insertion with 1 becomes $M^{\sigma_1}_{\sigma_3} 1_{\sigma_2}$
where $1_{\sigma_2} = 1$ and
$$
M^{\sigma_1}_{\sigma_3} = \sum_{\tau\in S_2} Wg(\sigma_1^{-1}\tau,d)\langle\langle\sigma_3|\tau\rangle\rangle
$$
with value $M^I_I = M^S_S = (q^3-1)/q/(q^4-1)$, $M^I_S = M^S_I = (q-1)/(q^4-1)$. We may refer to this informally as the "measurement triangle'', but it breaks down into a product of $M^\sigma_\tau$, the two-index $M$ tensor, and the ones tensor $1^\sigma$ and geometrically is a dot on one node and a line between two nodes in our simplicial complex diagrams. 

With this new tensor in mind, let us return to computing $\langle D^2\rangle$. This quantity is like $\langle (\partial N)^2\rangle$ we considered before but with a single measurement present and without the gradient triangle and the Hamiltonian node at the final time boundary present. The only allowed configurations here, following our earlier arguments, are the uniform $I$-domain, which was the only configuration in $\langle D\rangle$, and those in which a single $S$ domain terminates at the measurement.

The identity configuration contributes $M^I_I = (q^3-1)/q/(q^4-1)$ to $\langle D^2\rangle$, which vanishes like $q^{-2}$ at large-$q$. Recalling that $\langle D\rangle = q^{-m}$ with $m=1$ here, we see that at large-$q$, the variance is 
\begin{equation}
    \mathrm{Var}(D) = \langle D^2\rangle-\langle D\rangle^2 = q^{-2}\left((1-1/q^3)/(1-1/q^4)-1\right) +\ldots = -q^{-5}\frac{1-1/q}{1-1/q^4}+\ldots
\end{equation}
where $\ldots$ refers to the $S$-domain contributions. Since the first quantity here is negative, the $S$ domain must contribute at least order $q^{-5}$ to compensate at large $q$. 

Now, at large $q$, the $S$-domain contributions become straightforward to delineate in powers of $q$. The largest contributing $S$-domain is the configuration which has just one $S$ node occurring exactly at the measurement location that contributes $J^I_{IS}J^I_{SI}M^S_I=(q/(q^2+1))^2(q-1)/(q^4-1)$ to $\mathrm{Var}(D)$. This also vanishes like $q^{-5}$ as $q\to\infty$. We now find
\begin{equation}
\mathrm{Var}(D) = \left(\frac{q}{q^2+1}\right)^2\frac{q-1}{q^4-1}-\frac{1}{q^2}\frac{q-1}{q^4-1} +\ldots = -\frac{2q^2+1}{q^2(q^2+1)^2}\frac{q-1}{q^4-1} +\ldots
\end{equation} 
which is again negative! Now our leading contribution decays like $-2/q^7$ as $q\to\infty$ and again we need another contribution to compensate. Rather than continue, we will stop here and simply bound $\text{Var}(D)$, finding $\text{Var}(D) = O(q^{-7})$.

Given $\langle D\rangle = q^{-m}$ and $\text{Var}(D) = O(q^{-7})$, we can finally assess the error
\begin{align}
    |\langle \delta C\rangle| &\leq \frac{1}{c\langle D\rangle}\frac{\sqrt{\text{Var}(D)}}{\langle D\rangle}+96c\langle D\rangle+\frac{\text{Var}(D)}{(1-c)^2\langle D\rangle^3}\\
    &\leq \frac{1}{c}q^{2m-7/2}+96cq^{-m} + \frac{1}{(1-c)^2}q^{3m-14}
\end{align}
As a result, the error decays to zero in the large-$q$ limit provided $2m < 7/2$ and $3m < 14$ or it decays to zero for $m=1$. 

\begin{theorem}[Cost function is approximated by the leading term at large-$q$]\label{thm:cost_function}
The cost function $C(\theta) = \frac{N}{\langle D\rangle + R}$ can be expressed as a convergent series in $R$ and is given by the first term in the series asympotically as $q\to\infty$. 
\end{theorem}
\begin{proof}
Since the above results show $\langle \delta C\rangle)$ vanishes as $q\to \infty$, the expansion converges and the leading term dominates.
\end{proof}

Given this theorem, we keep the leading term in the series and find $C(\theta) = q^{m}N(\theta)$, and so we apply conjecture \cref{conj:ave_variance} and make statements about the behavior of $q^{2m}\langle (\partial N)^2\rangle$. interpreting the results as the behavior of $\mathrm{Var}(\partial C)$. Namely, that, aside from the factor of $q^{2m}$, the leading effect of measurements are to replace a $J$-triangle with a dot on one node and an $M$-line between two nodes in our simplicial complex diagrams.

We can divide the effect of measurements into two regimes: measurements arising close to the gradient triangle and measurements arising farther from it. Measurements arising close to the gradient enable an $S$-node between the gradient and the measurement location. These $S$-nodes, as before, come with many canceling contributions and the situation is complex. Hence, at the level of the present analysis, we can't rigorously determine the effect of these measurements.
\begin{definition}[Dilute Measurements]
    A system experiences dilute measurements if the number of measurements remains fix as the system size $N$ and or time scale $T$ go to infinity. 
\end{definition}

\begin{conjecture}[Irrelevance of measurements close to the gradient triangle]\label{conj:near_measurements}
    The large-$q$ contribution to the cost function in the presence of dilute measurements near the gradient triangle remains proportional to $q^{-N}$ except if two post-selected measurements sit exactly in front of a 2-qubit unitary where the gradient would be exactly zero.
\end{conjecture}
\begin{evidence}
    A measurement outside the lightcone of the gradient triangle is innocuous. Here the leading contribution is exactly the same as in the absence of measurements, a barren plateau scaling like $q^{-N}$. A measurement placed in the lightcone of the gradient triangle a distance $\ell$ away, with $\ell \ll N,T$, has a dominant configuration of an $S$-node, a line of $S$ configurations spanning from the gradient to the measurement. The cost of this $S$-node at large-$N$ is $4q^{-3-2\ell}$ and doesn't scale with $N$. It appears to completely mitigate the Barren plateau. But there is a second configuration with a cost $-4q^{-3-2\ell}$ which exactly cancels this cost and subdominant configurations arise. We have both a) seen other situations like the computation of $\text{Var}(D)$ above where such cancellations continue to happen order by order in $1/q$ and b) we know entanglement is extremely difficult to contain, with strong evidence from the MIPT where a finite density of measurements is necessary to cause a volume-law phase to become an area-law entagnlement phase. Hence the cancellations are likely associated with the spread of entanglement and so likely continue to arise for all configurations until the sideways-V shaped configuration with no canceling partner is found to dominate the variance of the gradient 
\end{evidence}

Measurements occurring close to the gradient triangle take part in the battle against the growth of entanglement, so this is the most interesting regime and worth further study. Here measurements try to reduce entanglement while scrambling produced by the random two-qubit unitary gates tries to grow entanglement. This battle is precisely the subject of the field of MIPTs where in the volume law phase measurements are unable to contain the growth of entanglement, which is precisely present here in the dilute measurement regime.  Outside of the special case where two measurement take place right after a gradient taken on a 2-qubit unitary, where the measurements, when post selected, collapse the state irrespective of the parameter values of the unitary, the extensive studies of the MIPT provide strong evidence that entanglement will win in the dilute measurement limit. But these studies while related are different from VQA problems as we have argued in \cref{subsec:mutual_info_results}. So the effect of measurements close to the gradient triangle remains an interesting problem to solve. 

When measurement are far from the gradient, however, and we are in the $T\gg N$, large-$N$ regime, they either do nothing to the leading rotated V-shaped $S$-domain, or they can locally deviate the boundary between the $S$-domain and $I$-domain as in Fig. \ref{fig:statemechBP}(d) presented in the main text. Namely, 
\begin{theorem}[Irrelevance of measurements far from the gradient triangle]\label{thm:far_measurements}
    The large-$q$ contribution to the variance of the gradient in the presence of dilute measurements far from the gradient triangle remains proportional to $q^{-N}$ in $q^{m}\text{Var}(\partial N)$. 
\end{theorem}
\begin{proof}
    Recall \cref{thm:BPatlargeT}. If measurements occur far from the gradient, farther than the size of the system $N$, then the sideways V-shaped configuration dominates. In this regime, there are two cases, one where the measurements take place away from a domain wall and one where they take place on the domain wall. When they are away from the domain wall, they introduce a factor of $M^S_S = M^I_I = (q^3-1)/(q^4-1)\sim 1/q$ as $q\to\infty$ and there is no benefit for moving the domain wall. Hence these measurements are innocuous. In the second case, where they land on the domain wall, they may cause the domain wall to change shape. But a measurement is local and the domain wall is global, it needs to span from the first qubit along the diagonal to the Hamiltonian node and then down to the last qubit along another diagonal. A local change to the domain wall cannot change the over all scaling $q^{-N}$ as $N\to\infty$. So dilute measurements in both the bulk and domain wall measurements fail to affect the barren plateau.
\end{proof}

\newpage
\section{Finite-size scaling analysis of measurement induced criticality}
\begin{figure*}[th]
\includegraphics[width=\textwidth]{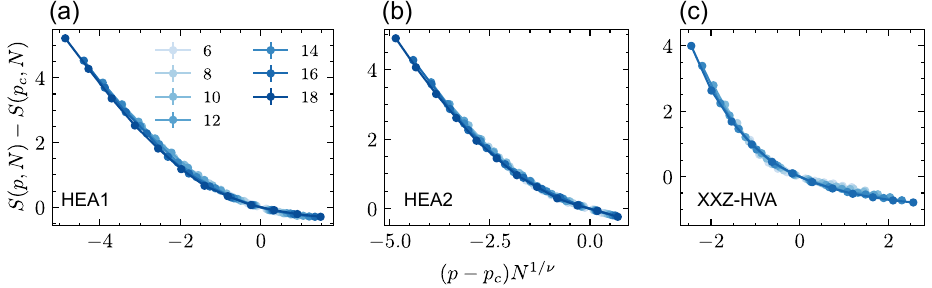}
\caption{\textbf{Finite-size scaling data collapse of the average entanglement entropy.} Entanglement dynamics starting from a computational basis state $\ket{00\cdots 0}$ was computed for \textbf{a} HEA1, \textbf{b} HEA2, and \textbf{c} XXZ-HVA ansatz used to extract the critical probability $p_c^{\mathrm{MIPT}}$, and scaling exponent $\nu$. XXZ-HVA lacks $N=18$ datapoints.}.
\label{fig::scaling_analysis}
\end{figure*}
\label{app:finite_size_scaling}
Although phase transitions are defined in the thermodynamic limit, we can use finite-size scaling analysis to study them numerically. With our exact state vector calculations, we were able to simulate systems only up to 18 qubits with $10^3$ samples for $N\leq 16$ and $500$ samples for $N=18$. We used the quality of the data collapse to determine the critical probability $p_c^{\mathrm{MIPT}}$, and critical exponent $\nu$. 

The model for the finite-size scaling collapse is given in \cref{eqn::scaling collapse}. If there is a continuous phase transition at $p=p_c^{\mathrm{MIPT}}$, then the left-hand side of this equation, computed close to $p_c^{\mathrm{MIPT}}$ should collapse into the scaling function of the parameter $(p-p_c^{\mathrm{MIPT}})N^{1/\nu}$. The fitting parameters are determined by the method described in Ref.  \cite{Skinner_MIPT_2019}. We discuss this briefly for completeness.

We define a cost function $R(p_c^{\mathrm{MIPT}}, \nu)$ for given values of critical parameters $p_c^{\mathrm{MIPT}}$ and $\nu$ which will be minimized to get the best estimates. $S(p_c^{\mathrm{MIPT}},N)$ for each system size $N$ is determined by linear interpolation and $x = (p-p_c^{\mathrm{MIPT}})N^{1/\nu}$, $y_N = S(p,N)- S(p_c^{\mathrm{MIPT}},N)$  is calculated for each data point. A family of curves $y_L(x)$ is obtained and the average of these curves calculated to get the mean $\bar{y}(x)$, to which all our data points will hopefully collapse. The cost function is defined as
\begin{equation}
    R = \sum_{i,L}\left[ y_L(x_i) - \bar{y}(x_i)\right]^2.
\end{equation}
The fitting cost function was minimized using L-BFGS-B routine available in SciPy library to get an estimate for $p_c^{\mathrm{MIPT}}$ and $\nu$. Once a good estimate was found, random initial values in their neighborhood were used as initial parameters and $5$ different estimates were collected. The values reported in \cref{tab:critical_parameters} are the means of those estimates and the error bars are the standard deviations. Better estimates for $N \rightarrow \infty $ can be found by iterating through the process of limiting the dataset to some $N$ and doing a fit for $p_c^{\mathrm{MIPT}}$ and $\nu$ vs $1/N$ and taking the y-intercept. We did not go through this exercise since the measurement-induced entanglement phase transition is not the main topic of our study.

\newpage
\section{Optimization traces for XXZ cost function}
\label{app:xxz_traces}
Here, we examine the optimization traces for a $XXZ$ Hamiltonian with the same variational circuits. Similar to the $Z_0 Z_1$ case of \cref{fig:z0z1_optimization}, we find that lower energy can be reached with a high likelihood with hybrid variational ansatzes compared to the unitary ansatzes. Before getting stuck in a quantum Zeno state, the circuits with a low probability of measurement ($p<0.2$) explore a broader range of energy as indicated by our variance results. However, the global minimum is never reached in this case because we chose the gapless region($\Delta<1$) of the phase diagram for optimization. Preparing a gapless ground state with either adiabatic evolution or VQE is a considerably tricky problem compared to a gapped one.

\begin{figure*}[th]
\includegraphics[width=\textwidth]{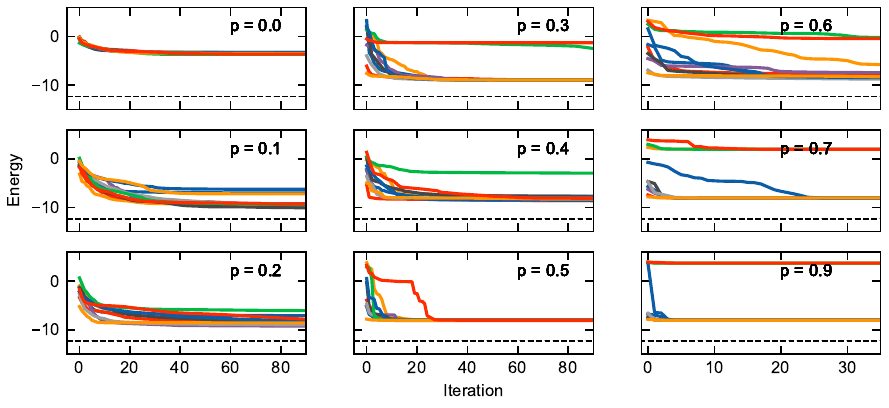}
    \caption{\textbf{Optimization with measurement and post-selection.}  Optimization traces are shown for an 8-qubit $XXZ$ Hamiltonian with anisotropy parameter $\Delta=0.5$. Each sub-figure corresponding to a measurement probability between 0 and 0.8 contains traces from 10 individual optimization runs carried out for the HEA2 ansatz with depth 20. The dashed line shows the exact ground state energy.}
    \label{fig:xxz_optimization}
\end{figure*}

\newpage

\newpage

\section{Determination of critical MILT transition probability}
\label{app:critical_point_analysis}

\begin{figure*}[th]
    \centering
    \includegraphics[width=0.8\textwidth]{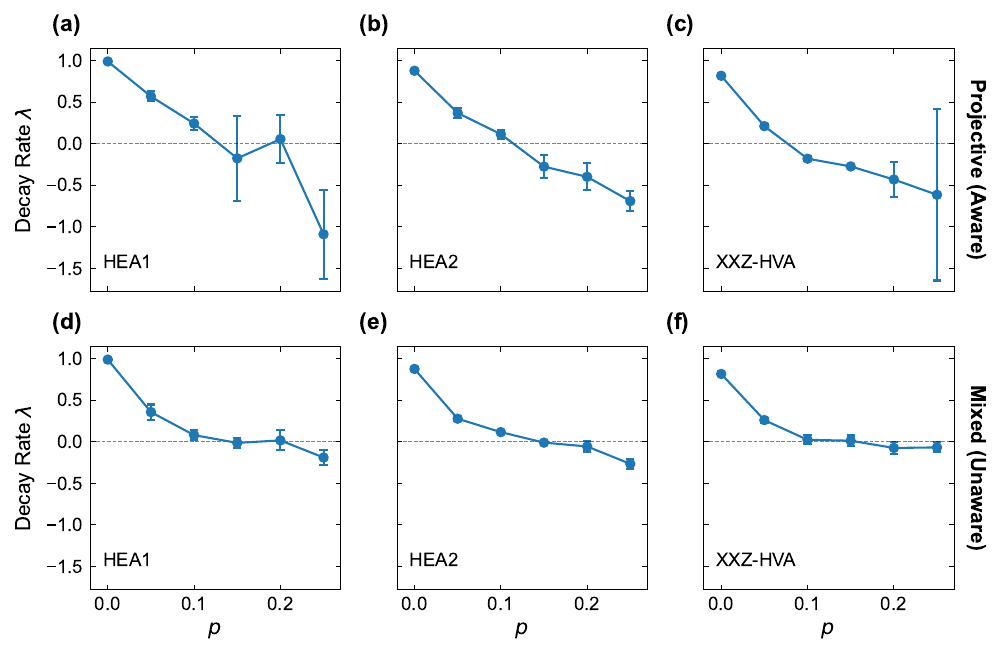}
    \caption{\textbf{BP decay rates versus measurement probability $p$.}The BP decay rate $\lambda(p)$ is plotted against the measurement probability $p$ for \textbf{(a, d)} HEA1, \textbf{(b, e)} HEA2, and \textbf{(c, f)} XXZ-HVA ansatzes. The top row \textbf{(a-c)} corresponds to the projective cost function, while the bottom row \textbf{(d-f)} corresponds to the mixed cost function. The decay rate $\lambda$ is extracted from the slope of $\log_2(\text{Var}) \propto -\lambda N$ for system sizes $N \in \{8, 10, 12, 14\}$. Error bars represent the standard error of the linear regression slope. The gray dashed line indicates $\lambda=0$, representing the absence of a BP. We identify the critical point $p_c^{\text{MILT}}$ as the onset of the non-decaying phase ($\lambda \approx 0$). This transition occurs at $p \approx 0.15$ for the HEA ansatzes and slightly earlier at $p \approx 0.10$ for the XXZ-HVA ansatz, both of which are significantly lower than the corresponding entanglement transition thresholds ($p_c^{\text{MIPT}}$).}
    \label{fig:decay_rate_analysis}
\end{figure*}

The BP, by definition, is the variance of the gradient of an observable decreasing exponentially with system size \cite{Sim_expressibility_2019,Haug_2021_expressibility_quantum_geometry, Holmes_expressibility_2022}, which can be represented by a scaling relationship $\text{Var}_{\boldsymbol{\theta}}(\partial_k C(\boldsymbol{\theta})) \propto e^{-\lambda N}$ . The collapse of these BPs can be qualitatively observed in the raw variance data around $p \approx 0.10\text{--}0.25$ in \cref{fig:projective_gradients}. Here, we undergo a procedure to more quantitatively determine this probability $p=p_c^{\text{MILT}}$.

For each measurement probability $p$, we performed a linear regression of the logarithmic variance, $\log_2(\text{Var}(\partial_k C(\boldsymbol{\theta}))) = -\lambda(p) N + C$, across system sizes $N \in \{8,10,12,14\}$. The slope of this fit yields the decay rate $\lambda(p)$, where $\lambda > 0$ indicates the presence of a BP. The validity of the exponential scaling ansatz was confirmed by the linear fits, which yielded a consistently high coefficient of determination ($R^2 > 0.90$) in the BP regime, confirming that the exponential model captures the system size dependence. Closer to the critical point, the $R^2$ values naturally decrease as the variance becomes effectively independent of the system size $N$, and the signal is lost underneath the variance of the stochastic sampling.

We define the critical point $p_c^{\text{MILT}}$ as the smallest probability $p$ at which the decay rate $\lambda(p)$ becomes statistically indistinguishable from zero (i.e., the onset of the non-decaying phase). Quantitatively, we observe that the critical measurement probability for the landscape transition ($p_c^{\text{MILT}}$) exhibits ansatz-dependence while remaining consistently bounded below the entanglement transition threshold ($p_c^{\text{MIPT}}$). For the hardware-efficient ansatzes (HEA1 and HEA2), the BP decay rate $\lambda(p)$ vanishes within statistical error at approximately $p \approx 0.15$ for both projective and mixed cost functions (see \cref{fig:decay_rate_analysis}a,b,d,e). In contrast, the XXZ-HVA ansatz displays a slightly earlier transition, with the decay rate becoming indistinguishable from zero at $p \approx 0.10$ (\cref{fig:decay_rate_analysis}c,f).

This variation suggests that the precise location of the landscape transition is influenced by the specific circuit structure and gate connectivity. Crucially, however, in all cases the MILT occurs distinctly earlier than the corresponding MIPT in \cref{tab:critical_parameters}, reinforcing our hypothesis that the vanishing of gradients is a distinct phenomenon from the volume-law to area-law entanglement transition.

\end{document}